\documentclass[floatfix, onecolumn, showpacs,nofootinbib]{revtex4-1}
\usepackage{natbib}
\usepackage{times}
\usepackage{amssymb,amsbsy,amsmath,amsfonts}
\usepackage{graphicx}
\usepackage{float}
\usepackage{morefloats}
\usepackage{rotating}
\usepackage{srcltx}
\usepackage{slashed}
\usepackage{subfigure}
\usepackage{hyperref}
\usepackage{color}
\begin{document}

\title{Recent developments in SU(3) covariant baryon chiral perturbation theory}

\author{ Lisheng Geng}

\affiliation{
School of Physics and Nuclear Energy Engineering, Beihang University,  Beijing 100191,  China\\
International Research Center for Nuclei and Partciles in the Cosmos, Beihang University, Beijing 100191, China
}

 \begin{abstract}
Baryon chiral perturbation theory (BChPT), as an effective field theory of low-energy quantum chromodynamics (QCD),
has played and is still playing an important role in our understanding of non-perturbative strong-interaction phenomena.
In the past two decades, inspired by the rapid progress in lattice QCD simulations and the new experimental campaign to
study the strangeness sector of  low-energy QCD, many efforts have been made to develop a fully covariant BChPT and to test its validity in all scenarios. These new endeavours
have not only deepened our understanding of some long-standing problems, such as the power-counting-breaking problem and the convergence problem, but also resulted in theoretical tools that can be confidently applied
to make robust predictions. Particularly, the manifestly covariant BChPT supplemented with the extended-on-mass-shell (EOMS) renormalization scheme has been shown to satisfy all analyticity and symmetry constraints and  converge
relatively faster compared to its non-relativistic and infrared counterparts.
In this article, we provide
a brief review of the fully covariant  BChPT and its
latest applications in the $u$, $d$, and $s$ three-flavor sector.

\end{abstract}


\date{\today}

\maketitle

\section{Introduction}
Four fundamental interactions have so far been identified in nature: electromagnetism,  weak interaction, strong interaction, and gravitation.
 Among them, the strong interaction is widely acknowledged as one of
the most difficult to comprehend. In particular, the low-energy or non-perturbative strong-interaction phenomena have afflicted physicists for several decades. The non-abelian quantum field theory, generally accepted underlying the strong interaction, is the quantum chromodynamics (QCD) and was proposed about 30 years ago. However, two peculiar features of QCD make its solution in the low-energy region difficult. First, the running coupling constant, $\alpha_s$, becomes small as the energy (or four momentum) transfer increases--a phenomenon called the asymptotic freedom~\cite{Gross:1973id,Politzer:1973fx}, which implies that the running coupling constant becomes large at low energies.The large running coupling constant renders a perturbative treatment of QCD useless. Another closely related feature is color confinement. That is, although the degrees of freedom in QCD are quarks and gluons, which carry a quantum-number color, in nature only color-neutral objects, mesons and baryons, are observed. This leads to the complication that in QCD the underlying degrees of freedom and the observables are different. As a result, in studies of the low-energy strong-interaction phenomena, despite the fact that there exists a fundamental theory, one has to turn to effective field theories (EFTs) or models, which are motivated by the various symmetries of QCD, or
brute-force numerical methods, such as lattice quantum chromodynamics (LQCD)~\cite{Wilson:1974sk}.

 One of such EFTs is the chiral perturbation theory (ChPT).
 In the massless limit, the QCD Lagrangian is invariant under the separate transformation of left and right handed quark fields, i.e.,
$\mathcal{L}^0_\mathrm{QCD}$ is invariant under $\mathrm{SU}(N)_L
\times\mathrm{SU}(N)_R$ ($N=2$ or 3), which is termed as chiral symmetry. On the other hand, chiral symmetry is spontaneously broken, resulting in the appearance of eight massless Nambu-Goldstone bosons (NGBs). Experimentally the lowest-lying pseudoscalar octet can be identified as the corresponding NGBs. Furthermore,
 the masses of the $u$, $d$ quarks and, to a lesser extent, that of the strange quark are small enough so that a perturbative treatment of their masses is appropriate~\cite{Donoghue:1992dd}.

 It was Weinberg who first proposed that
the soft-pion results of current algebra can be recast into the language of an EFT~\cite{Weinberg:1978kz}. It was then systematically developed in the mesonic sector with two ~\cite{Gasser:1983yg} and three flavors~\cite{Gasser:1984gg}. Its application to the one-baryon sector came a few years later and met an unexpected difficulty, i.e., the combination of the power-counting (PC) rule of Weinberg and the modified minimal subtraction scheme ($\overline{\mathrm{MS}}$) yields the so-called power-counting-breaking (PCB) terms~\cite{Gasser:1987rb}, which
invalidates a systematic chiral expansion. To remove these PCB terms and restore the proper chiral power-counting rule,
several approaches have been proposed. The first and most extensively studied is the heavy-baryon (HB) ChPT~\cite{Jenkins:1990jv}. It has been successfully applied to study a variety of physical phenomena, particularly in the $u$ and $d$ two-flavor  sector (see, e.g., Ref.~\cite{Bernard:1995dp} for a review of early applications). However, from the beginning, the drawbacks and limitations of the heavy-baryon ChPT are  well known. For instance, extra care has to be paid to make sure that reparameterisation invariance~\cite{Luke:1992cs} or Lorentz invariance~\cite{Ecker:1995rk} is satisfied by the chiral Lagrangians of the heavy-baryon ChPT  by keeping track of various $1/m$ corrections, where $m$ is the heavy baryon mass. As a result, the Lagrangians of the heavy-baryon ChPT usually have more terms that its relativistic counterparts at a given order~\cite{Ecker:1995rk} and analysis of loop amplitudes can become non-trivial, even in the simple case of wave-function renormalization~\cite{Ecker:1997dn}. Furthermore, the perturbation series in the heavy-baryon ChPT may fail to converge in part of the low-energy region, because in the infinite heavy baryon mass limit the analytical structure (poles and cuts) of certain amplitudes may be disturbed, e.g., the Dirac, Pauli, and scalar form factors of the nucleon at $t=4 m_\pi^2$ ~\cite{Bernard:1995dp}.

Over the years, a number of relativistic formulations of BChPT were proposed by different authors~\cite{Tang:1996ca,Becher:1999he,Fuchs:2003qc}. Among these, the infrared (IR)~\cite{Becher:1999he} and extended-on-mass-shell (EOMS)~\cite{Fuchs:2003qc} schemes  are the mostly studied. The infrared approach restores the power-counting rule in such a way that analyticity is broken at a scale of twice the baryon mass, which was originally believed to be of minor importance~\cite{Becher:1999he}. However, later studies found that the effects start to show up at much smaller meson masses, which are relevant for studies of the
light-quark mass evolution of certain observables, such as the nucleon magnetic moment~\cite{Pascalutsa:2004ga}, and for studies performed in the three-flavor sector~\cite{Geng:2008mf}. The EOMS approach, on the other hand, is free of such problems and appears to be a better formulation of covariant BChPT. Apart from the fact that a fully covariant formulation of BChPT is formally more appealing, it has been
demonstrated that the EOMS approach converges relatively faster, a very important and practical feature for an EFT (see, e.g., Refs.~\cite{Geng:2008mf,MartinCamalich:2010fp,Geng:2010df}).

In the past decade, lattice QCD has developed into
an indispensable tool to study low-energy strong-interaction phenomena (for some recent reviews, see, e.g., Refs.~\cite{Hagler:2009ni,Bazavov:2009bb,Colangelo:2010et,Beane:2010em,Aoki:2012tk,Fodor:2012gf,Durr:2013qk}).
To reduce computing cost, most LQCD simulations employed larger than physical light-quark masses,~\footnote{See, however, Refs.~\cite{Durr:2010aw,Durr:2010vn,Bazavov:2012xda} for a few recent simulations performed at the physical point.} finite volume and lattice spacing. As a result, to obtain the physical value of any observable
simulated in the lattice, extrapolations to the physical world are necessary. ChPT provides an useful framework to perform such extrapolations and to estimate the induced uncertaintiesá~\cite{Golterman:2009kw}. On the other hand,
the LQCD quark-mass, volume, and lattice-spacing dependent results also offer a unique opportunity to help determine the many unknown low-energy constants (LECs) of ChPT~\cite{Ren:2012aj}.  In many recent studies, it has
been shown that the EOMS ChPT can provide a better description of the LQCD quark-mass dependent results than  its non-relativistic counterpart~\cite{MartinCamalich:2010fp,Geng:2010df}.

This paper is organized as follows. In Section II, we collect the relevant chiral Lagrangians needed for later discussions,
introduce the power-counting-breaking problem encountered in constructing a covariant BChPT, explain briefly the heavy-baryon, infrared, and
extended-on-mass-shell formulations, and point out
their advantages and limitations.
In Section III, we show a few recent applications of the EOMS BChPT in the $u$, $d$ and $s$ three-flavor sector,
including the magnetic moments of the octet and decuplet baryons, the masses and sigma terms of the octet baryons, and the hyperon vector couplings. 
The recent developments in the $u$ and $d$ two-flavor sector and the extension to
the heavy-light system are covered in Sections IV and V, respectively.
Section VI contains a brief summary and outlook.

\section{Covariant baryon chiral perturbation theory}
In this section, we briefly explain the heavy-baryon, the infrared and the extended-on-mass-shell formulations of baryon chiral perturbation theory (BChPT).
This article is intended to be neither a pedagogical introduction to ChPT nor an exhaustive collection of all the recent works.
For both purposes, there exist excellent monographs and review articles~\cite{Meissner:1993ah,Ecker:1994gg, Bernard:1995dp,Pich:1995bw,Bijnens:2006zp,Bernard:2007zu,Scherer:2009bt,Scherer:2012zzd}. Furthermore, we concentrate our discussions on
the one-baryon sector in the three-flavor space.

The chiral Lagrangians in the one-baryon sector can be generically written as
\begin{equation}
\mathcal{L}=\mathcal{L}_{MB}+\mathcal{L}_M,
\end{equation}
where $\mathcal{L}_{MB}$ contains the part of the Lagrangians describing the interaction of the low-lying baryons with the NGBs while
$\mathcal{L}_M$ describes the self-interaction of the NGBs. In general, the
$\mathcal{L}_{MB}$ and $\mathcal{L}_M$ contain a series of terms ordered by the so-called chiral power counting, i.e.,
\begin{equation}\label{eq:blag}
\mathcal{L}_{MB}=\mathcal{L}_{MB}^{(1)}+\mathcal{L}_{MB}^{(2)}+\mathcal{L}_{MB}^{(3)}+\cdots,
\end{equation}
\begin{equation}\label{eq:mlag}
\mathcal{L}_M=\mathcal{L}_M^{(2)}+\mathcal{L}_M^{(4)}+\mathcal{L}_M^{(6)}+\cdots.
\end{equation}
The chiral order, indicated by the superscript, is assigned in the following way:
the mass of a NGB and a derivative on its field are counted as of $\mathcal{O}(p)$. The mass of a baryon is
counted as of $\mathcal{O}(p^0)$ and the same is for the derivative on its field. However, $\slashed{P}-m_B$, with $P^\mu$ the baryon four momentum, is
counted as of $\mathcal{O}(p)$. As a result, the baryon propagator $\frac{1}{\slashed{P}-m_B}$ is counted of as $O(p^{-1})$ and the meson propagator
$\frac{1}{P^2-m_M^2}$ as $\mathcal{O}(p^{-2})$, with $p$ denoting a generic small quantity, $m_B$ and $m_M$ the masses of
the baryon and the meson. For a Feynman diagram consisting of
$L$ loops,
$N_B$ baryon propagators, $N_M$ NGB propagators, and $m$ $k$-th order vertices, its chiral order $n$ is defined as
\begin{equation}\label{eq:pcb}
n=4 L- N_B-2N_M+\sum_ {m,k} m k.
\end{equation}
With the chiral power-counting rule defined above and together with the $\overline{\mathrm{MS}}$ scheme,
this completes the definition of a power-counting rule for ChPT in the mesonic sector, i.e., in studies of the self-interaction of
the NGBs.

Once a matter field, such as the baryon field discussed here, is introduced, the above power-counting rule
is violated. That is to say, lower-order analytical terms appear in a nominally higher-order calculation. For instance,
in the calculation of the nucleon self-energy at $\mathcal{O}(p^3)$, one finds $\mathcal{O}(p^0)$ and $\mathcal{O}(p^2)$ terms. Although these
terms do not change the underlying physics, they have two unwelcome consequences for an EFT. First, they can completely change the natural values of the lower-order LECs. Second, they render an order-by-order analysis difficult.  Both complicate a lot the study of baryon properties. In the following, we will explain in detail how
one can recover the proper power-counting rule defined in Eq.~(\ref{eq:pcb}).

A superficial difference between the Lagrangians (\ref{eq:blag}) and (\ref{eq:mlag}) is that
Lagrangian (\ref{eq:mlag}) contains only even-order terms while Lagrangian (\ref{eq:blag}) has both even- and odd-order terms
starting from order 1. It has a direct consequence: the BChPT converges slower than the meson ChPT.
Furthermore, there are more LECs in the baryon ChPT than in the meson ChPT at a given order. For instance, in the mesonic sector, there are two LECs at $\mathcal{O}(p^2)$ and 12 LECs at $\mathcal{O}(p^4))$~\cite{Gasser:1984gg} while in the one-baryon sector,
the corresponding numbers of LECs are three for $\mathcal{O}(p)$ and 16 for $\mathcal{O}(p^2)$ ~\cite{Oller:2006yh}, respectively.~\footnote{
In this work, unless stated otherwise, we refer to ChPT in the three-flavor sector.}

The leading order and next-to-leading order meson Lagrangians are:~\footnote{The complete next-to-next-to-leading order ($O(p^6)$) Lagrangians can be found in Ref.~\cite{Bijnens:1999sh}.}
\begin{equation}\label{eq:blag1}
\mathcal{L}_M^{(2)}=\frac{F_0^2}{4}\mathrm{Tr}[D_\mu U(D^\mu U)^\dagger]+\frac{F_0^2}{4}\mathrm{Tr}[\chi U^\dagger+U\chi^\dagger],
\end{equation}
\begin{eqnarray}
\mathcal{L}_M^{(4)}&=&L_1\left\{\mathrm{Tr}[D_\mu U (D^\mu U)^\dagger]\right\}^2+L_2\mathrm{Tr}[D_\mu U(D_\nu U)^\dagger]\mathrm{Tr}[
D^\mu U(D^\nu U)^\dagger]\nonumber\\
&&+L_3\mathrm{Tr}[D_\mu U (D^\mu U)^\dagger D_\nu U (D^\nu U)^\dagger]+
L_4\mathrm{Tr}[D_\mu U (D^\mu U)^\dagger]\mathrm{Tr}[\chi U^\dagger+U\chi^\dagger]\nonumber\\
&&+L_5 \mathrm{Tr}[D_\mu U (D^\mu U)^\dagger (\chi U^\dagger+U\chi^\dagger)]+L_6\{\mathrm{Tr}[\chi U^\dagger+ U\chi^\dagger]\}^2\nonumber\\
&&+L_7\{\mathrm{Tr}[\chi U^\dagger-U\chi^\dagger]\}^2+L_8\mathrm{Tr}[U\chi^\dagger U \chi^\dagger+\chi U^\dagger \chi U^\dagger]\nonumber\\
&&-i L_9 \mathrm{Tr}[f^R_{\mu\nu} D^\mu U (D^\nu U)^\dagger+f^L_{\mu\nu}(D^\mu U)^\dagger D^\nu U]+L_{10}\mathrm{Tr}[U f^L_{\mu\nu} U^\dagger f_R^{\mu\nu}]\nonumber\\
&&+H_1\mathrm{Tr}[f^R_{\mu\nu}f_R^{\mu\nu}+f^L_{\mu\nu}f_L^{\mu\nu}]+H_2\mathrm{Tr}[\chi\chi^\dagger],
\end{eqnarray}
where $U=\exp\left(i\frac{\Phi}{F_0}\right)$, $D_\mu U =\partial_\mu U-i r_\mu U + i U l_\mu$,
$f^R_{\mu\nu}=\partial_\mu r_\nu-\partial_\nu r_\mu-i[r_\mu,r_\nu]$, $f^L_{\mu\nu}=\partial_\mu l_\nu-\partial_\nu l_\mu -i [l_\mu,l_\nu]$,
$\chi=2 B_0 (s+ i p)$, and
$\Phi$ collects the NGB fields
\begin{equation}
\Phi=\left(\begin{array}{ccc}
\frac{1}{\sqrt{3}}\eta+\pi^0 & \sqrt{2}\pi^+ & \sqrt{2}K^+\\
\sqrt{2}\pi^-&\frac{1}{\sqrt{3}}\eta-\pi^0 &\sqrt{2} K^0\\
\sqrt{2}K^- &\sqrt{2}\bar{K}^0 &-\frac{2}{\sqrt{3}}\eta
\end{array}\right).
\end{equation}
In the above Lagrangians, $F_0$ is the pseudoscalar meson decay constant in the chiral limit, $B_0=-\langle0| q\bar{q}|0\rangle/F_0^2$ with $\langle0| q\bar{q}|0\rangle$ the SU(3) quark condensate,
$s$, $p$, $r_\nu=v_\nu+a_\nu$, $l_\nu=v_\nu-a_\nu$ are the external scalar, pseudo scalar, right-handed, and left-handed currents with $v_\nu$ and $a_\nu$ the external vector and axial-vector currents.
Presently, the LECs can be determined using either empirical inputs,  estimates based on the resonance saturation assumption, or lattice QCD. See Ref.~\cite{Bijnens:2011tb}
for a latest review on the present status in the mesonic sector.

In the one-baryon sector, the leading order Lagrangian has the following  form:
\begin{equation}\label{eq:MBLO}
\mathcal{L}_{MB}^{(1)}=\langle\bar{B}(i\slashed{D} -m_0) B\rangle +\frac{F}{2}\langle\bar{B}\gamma^\mu\gamma_5[u_\mu,B]\rangle+\frac{D}{2}\langle\bar{B}\gamma^\mu\gamma_5 \{u_\mu,B\}\rangle,
\end{equation}
where $u_\mu=i\{u^\dagger (\partial_\mu-i r_\mu)u-u(\partial_\mu-i l_\mu) u^\dagger\}$ with $u=\sqrt{U}$.  It contains three LECs, $m_0$, $D$, and $F$. At next-to-leading order, the number of LECs increases to 16~\cite{Oller:2006yh}:
\begin{eqnarray}
\mathcal{L}_{MB}^{(2)}&=&b_D\langle \bar{B}\{\chi_+,B\}\rangle+b_F\langle \bar{B}[\chi_+,B]\rangle+b_0\langle\bar{B}B\rangle\langle\chi_+\rangle\nonumber\\
&&+b_1\langle\bar{B}[u^\mu,[u_\mu,B]]\rangle+b_2\langle\bar{B}\{u^\mu,\{u_\mu,B\}\}\rangle\nonumber\\
&&+b_3\langle\bar{B}\{u^\mu,[u_\mu,B]\}\rangle+b_4\langle\bar{B}B\rangle\langle u^\mu u_\mu\rangle\nonumber\\
&&+ib_5\left(\langle \bar{B}[u^\mu,[u^\nu,\gamma_\mu D_\nu B]]\rangle-\langle B\overleftarrow{D}_\nu[u^\nu,[u^\mu,\gamma_\mu B]]\rangle\right)\nonumber\\
&&+ib_6\left(\langle \bar{B}[u^\mu,\{u^\nu,\gamma_\mu D_\nu B\}]\rangle-\langle B\overleftarrow{D}_\nu\{u^\nu,[u^\mu,\gamma_\mu B]\}\rangle\right)\nonumber\\
&&+ib_7\left(\langle \bar{B}\{u^\mu,\{u^\nu,\gamma_\mu D_\nu B\}\}\rangle-\langle B\overleftarrow{D}_\nu\{u^\nu,\{u^\mu,\gamma_\mu B\}\}\rangle\right)\nonumber\\
&&+ib_8\left(\langle\bar{B}\gamma_\mu D_\nu B\rangle-\langle B\overleftarrow{D}_\nu\gamma_\mu B\rangle \right)\langle u^\mu u^\nu\rangle
+id_1\langle\bar{B}\{[u^\mu,u^\nu],\sigma_{\mu\nu} B\}\rangle\nonumber\\
&&+id_2\langle\bar{B}[[u^\mu,u^\nu],\sigma_{\mu\nu} B]\rangle+id_3\langle\bar{B} u^\mu\rangle\langle u^\nu \sigma_{\mu\nu} B\rangle\nonumber\\
&&+d_4\langle\bar{B}\{f_+^{\mu\nu},\sigma_{\mu\nu}B\}\rangle+d_5\langle\bar{B}[f_+^{\mu\nu},\sigma_{\mu\nu}B]\rangle.
\end{eqnarray}
At the third order ($\mathcal{O}(p^3)$), the number of independent LECs increases to 78~\cite{Oller:2006yh,Frink:2006hx,Oller:2007qd}. In general, the number of LECs increases rapidly with each increasing order--a feature of non-renormalizable EFTs. Fortunately, not all the LECs contribute to
a particular physical process and therefore even at higher orders, BChPT retains its predictive power for certain observables, such as the hyperon vector form factors at zero four momentum transfer, where
no LECs appear up to $\mathcal{O}(p^4)$~\cite{Geng:2009ik}.
\subsection{Power-counting-breaking and its restoration}
\begin{figure}[t]
\centerline{\includegraphics[scale=0.4]{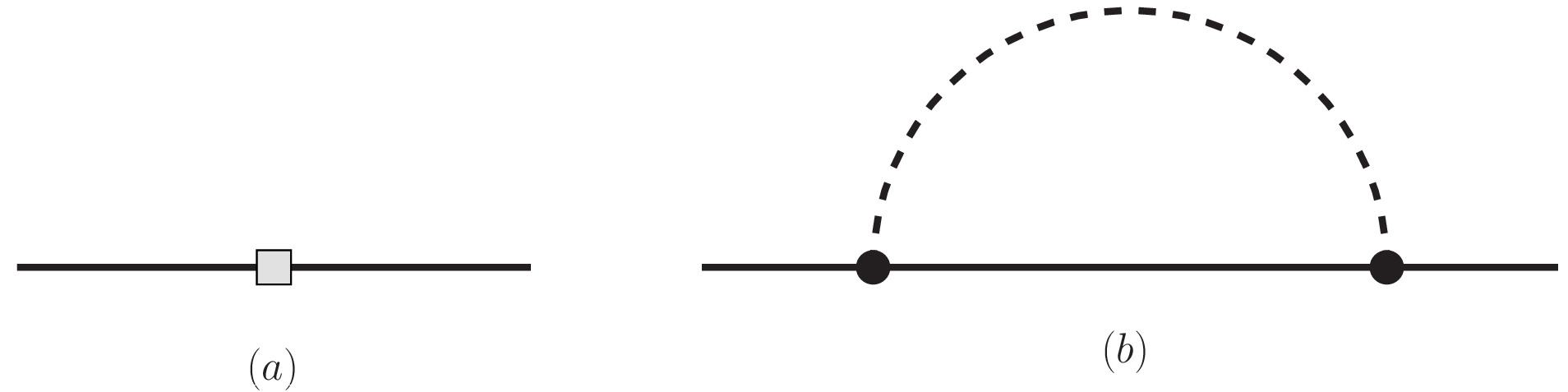}}
\caption{Feynman diagrams contributing to the nucleon mass at NNLO BChPT. Diagram (a) represents tree level contributions, both at order 0 and order 2, while diagram (b) represents contributions at NNLO (order 3). The solid line denotes the nucleon and the dashed line denotes a pseudoscalar meson. The solid dot denotes an order 1 vertex coming from the Lagrangian of Eq.~(\ref{eq:MBLO}).\label{fig:NNLOmass}}
\end{figure}

 The chiral power-counting rule of Eq.~(\ref{eq:pcb}) is known to be violated in the one-baryon sector~\cite{Gasser:1987rb}. In the following, we will use the nucleon mass
 as an example to demonstrate the power-counting-breaking problem. Up to NNLO, the nucleon mass receives contributions from both tree and loop diagrams (see Fig.~\ref{fig:NNLOmass}). Schematically, it reads as
 \begin{equation}\label{eq:mass3}
 m_N=m_0 + 2(b_0+2b_F)m_\pi^2 + \frac{3(D+F)^2}{64F_0^2\pi^2}\mathrm{loop}^{(3)}(m_\pi),
 \end{equation}
 where $\mathrm{loop}^{(3)}$ indicates the one-loop contribution from diagram (b) of
 Fig.~\ref{fig:NNLOmass}. According to the chiral power-counting rule of Eq.~(\ref{eq:pcb}), diagram (b) counts as of order 3.~\footnote{We limit our discussion to the two flavor case and neglect the contribution of the virtual $\Delta(1232)$. A complete SU(3) NNLO calculation
 in the EOMS BChPT can be found in
 Ref.~\cite{MartinCamalich:2010fp} .}A direct calculation of the contribution of diagram (1b)
 to the nucleon mass yields
 \begin{eqnarray}\label{eq:loop3}
 \mathrm{loop}^{(3)}&\propto&i\int\frac{d^4 k}{(2\pi)^4}\frac{\slashed{k}(\slashed{k}-\slashed{p}+m_0)\slashed{k}}{(k^2-m_M^2+i\epsilon)((p-k)^2-m_0^2+i\epsilon)}\\
 &=&\underline{ 2m_0^3\left[1+\log\left(\frac{\mu^2}{m_0^2}\right)\right]+2m_0 m_\pi^2\left[2+\log\left(\frac{\mu^2}{m_0^2}\right)\right]}
-\frac{1}{m_0}\left\{\log\left(\frac{m_\pi^2}{m_0^2}\right)m_\pi^4 +2m_\pi^3\sqrt{4m_0^2-m_\pi^2}\arccos\left(\frac{m_\pi}{2 m}\right)\right\}.\nonumber
 \end{eqnarray}
The above result is divergent and therefore in the second line we have utilized the $\overline{\mathrm{MS}}$ scheme to remove the divergence. Clearly, according to the chiral power-counting rule of Eq.~(\ref{eq:pcb}), the underlined terms are of order 0 and 2, respectively, and therefore do not satisfy the specified chiral order of this diagram, $\mathcal{O}(p^3)$.  These terms can be removed in a systematic way through
a number of different approaches, such as the heavy-baryon~\cite{Jenkins:1990jv}, the infrared~\cite{Becher:1999he}, and the extended-on-mass-shell~\cite{Fuchs:2003qc} schemes. The explicit expressions of  the nucleon mass up to $\mathcal{O}(p^3)$ in these three approaches can be found in Ref.~\cite{Pascalutsa:2011fp}. In the following, we briefly summarize the essential features of the heavy-baryon ChPT~\cite{Jenkins:1990jv}, infrared BChPT~\cite{Becher:1999he}, and extended-on-mass-shell BChPT~\cite{Fuchs:2003qc}.

\subsection{The heavy baryon (HB) approach}
In the heavy-baryon (HB) approach~\cite{Jenkins:1990jv}, one separates the nucleon four-momentum $p$ into a large piece and a soft residual component
\begin{equation}
p^\mu=m v^\mu+k_s^\mu.
\end{equation}
The velocity four-vector $v^\mu$ satisfies
\begin{equation}
v^2=1,\quad v^0\ge1,
\end{equation}
which can be taken to be $v^\mu=(1,0,0,0)$ for practical purposes. Note that with this choice of $v^\mu$,
 the soft residual momentum satisfies\ $k_s^0=-k_s^2/(2m)=E-m\ll m$.
 The essential idea is that although the zero-component of the nucleon four momentum
is not small, the soft residual component $k_s$ is small compared to the nucleon mass.

One can also decompose the relativistic nucleon field into two velocity-dependent fields
\begin{equation}
\psi(x)=e^{-imv\cdot x}(H_v(x)+L_v(x)),
\end{equation}
where $L_v=e^{+imv\cdot x}P_{v+}\psi$ and $H_v=e^{+imv\cdot x}P_{v-}\psi$ are
called light and heavy components, respectively, with the projection operators
\begin{equation}
P_{v\pm}=\frac{1}{2}(1+\slashed{v}).
\end{equation}
It can be easily shown that $\slashed{v}L_v=L_v$ and $\slashed{v}H_v=-H_v$. From the
relativistic Lagrangian of Eq.~(\ref{eq:MBLO}), employing either the equation of motion technique or
the path integral technique~(see, e.g., Ref.~\cite{Bernard:1992qa}), one can obtain the corresponding leading order heavy baryon Lagrangian
\begin{equation}
\widehat{\mathcal{L}}_{\pi N}^{(1)}=\bar{L}_v(iv\cdot D+g_A S_v \cdot u)L_v,
\end{equation}
where $\widehat{\hspace{0.3cm}}$ indicates that the Lagrangian is in the heavy-baryon formalism, $S^\mu_v\equiv\frac{i}{2}\gamma_5\sigma^{\mu\nu}v_\nu$ is the spin
matrix, and $g_A=D+F$. Note that the nucleon mass has disappeared from the leading order Lagrangian and only appears in terms of higher orders
 as powers of $1/m$. The heavy-baryon propagator, derived from the above Lagrangian, has the following form
\begin{equation}
G_v(k)=\frac{P_{v+}}{v\cdot k+i0^+}.
\end{equation}
The complete heavy-baryon Lagrangians up to and including order $p^4$ can be found in Ref.~\cite{Fettes:2000gb}.

Using the heavy-baryon Lagrangians and propagators together with the dimensional regularization scheme, $\widetilde{\mathrm{MS}}$, yields
the heavy-baryon ChPT. It has been widely applied to study various physical
observables and in most cases turned out to be successful. For a comprehensive review of early applications, see, e.g., Ref.~\cite{Bernard:1995dp}.

On the other hand,  it was noted in the very beginning that the $1/m$ expansion can create analyticity problems under specific
kinematics or, in other words, fails to converge in part of the low-energy region~\cite{Bernard:1995dp}. For instance, the nucleon scalar form factor calculated in the heavy-baryon ChPT explodes in the vicinity of $t=4m^2_\pi$~\cite{Bernard:1995dp,Becher:1999he}. In addition,
in the heavy-baryon ChPT one has to keep track of various $1/m$ corrections resulting from the non-relativistic
reduction of the relativistic Lagrangians and, therefore, the number of terms in the effective Lagrangians increases more rapidly than in its covariant counterparts.

To overcome the drawbacks of the heavy-baryon ChPT~\footnote{A  finite-range regularization (FRR) scheme has been proposed to
improve the convergence behaviour of the heavy-baryon ChPT~\cite{Stuckey:1996qr,Donoghue:1998bs,Young:2002cj,Young:2002ib,Leinweber:2003dg}.  It has
been rather successfully applied to study a variety of physical observables, particularly, in connection with LQCD simulations, their light-quark mass dependences (see, e.g., Refs.~\cite{Wang:2008vb,Young:2009zb,Wang:2010hp,Shanahan:2011su,Shanahan:2012wh,Wang:2012hj}). } and  derive a meaningful relativistic formulation of BChPT, many efforts have been taken in the past two decades. These efforts
have deepened our understanding of the power-counting-breaking problem and provided a number of useful solutions. Two most widely used formulations are the infrared (IR) ~\cite{Becher:1999he} and the extended-on-mass-shell (EOMS)~\cite{Fuchs:2003qc} schemes.
\subsection{The infrared (IR) scheme} \label{sec:ir}
The infrared scheme was introduced by Becher and Leutwyler~\cite{Becher:1999he}, following the approach of
Tang and Ellis~\cite{Tang:1996ca}. It  is based on the observation that in $D$ dimensions,
the infrared singular part of a loop diagram, which is free of PCB pieces, can
be separated from the regular part, which is a polynomial of the momenta and quark masses. The regular part
contains, in addition to the PCB terms, analytic higher-order pieces, which in principle can be absorbed by a redefinition of  the LECs.
To show the origin of the infrared singularities and how to remove the PCB terms in the IR scheme, we
use the following integral as an example
\begin{equation}\label{eq:example}
G=i \int\frac{d^D q}{(2\pi)^D}\frac{1}{[(P-q)^2-m_0^2+i\epsilon](q^2-m_\pi^2+i\epsilon)},
\end{equation}
where $D$ denotes the number of space-time dimensions, $m_0$ and $m_\pi$ are the chiral limit nucleon mass and the lowest-order pion mass. This integration can be easily performed by use of the Feynman parameterization method. Introducing a Feynman parameter $z$ and by use of the following identity,
\begin{equation}
\frac{1}{ab}=\int^1_0 \frac{dz}{[az+b(1-z)]^2},
\end{equation}
one obtains
\begin{equation}
G=i \int\frac{d^D q}{(2\pi)^D} \int^1_0 \frac{dz}{((P-zq)^2-\mathcal{M}^2)^2},
\end{equation}
where $\mathcal{M}^2=P^2z(z-1)+ m_0^2 z +  m_\pi^2(1-z) - i\epsilon$.
The integration over $q$ can be easily performed, yielding
\begin{equation}
G=-\frac{1}{(4\pi)^{D/2}}\Gamma(2-D/2)\int_0^1 dz (\mathcal{M}^2) ^{D/2-2} .
\end{equation}
 It was shown by Becher and Leutwyler that the integral $G$ can be divided into two parts, the infrared singular part $I$ and the
remaining regular part $R$, $ G=I+R$, defined as
\begin{equation}
I=-\frac{1}{(4\pi)^{D/2}}\Gamma(2-D/2)\int_0^\infty dz (\mathcal{M}^2) ^{D/2-2} ,
\end{equation}
\begin{equation}
R=\frac{1}{(4\pi)^{D/2}}\Gamma(2-D/2)\int_1^\infty dz(\mathcal{M}^2) ^{D/2-2} .
\end{equation}
It can be shown that for non-integer $D$ the integral $I$ is proportional to a non-integer power of the pion mass $\sim M^{D-3}$ and satisfies the power-counting rule of Eq.~(\ref{eq:pcb}).
On the other hand, the remaining regular part $R$ contains non-negative powers of the pion mass and violates the power-counting rule,
for arbitrary $D$~\cite{Becher:1999he}.  The IR scheme then dictates that one takes into account  only the infrared part and drops the regular part, with the argument that their contributions can be absorbed by
the corresponding LECs.

Calculating these three integrals explicitly in $D$ dimensions and then taking the limit of $D\rightarrow4$, one obtains
\begin{equation}\label{eq:GMSbar}
G=\frac{1}{16\pi^2}\left\{\underline{-2+\log\left(\frac{m_0^2}{\mu^2}\right)}+\frac{m_\pi^2}{m_0^2}\log\left(\frac{m_\pi}{m_0}\right)+\frac{2m_\pi}{m_0}\sqrt{1-\frac{m_\pi^2}{4m_0^2}}\arccos\left(\frac{m_\pi}{2m_0}\right)\right\},
\end{equation}
\begin{equation}
G_\mathrm{IR}\equiv I=\frac{1}{16\pi^2}\left\{ \frac{2m_\pi}{m_0}\left[\sqrt{1-\frac{m_\pi^2}{4m_0^2}}\arccos\left(-\frac{m_\pi}{2m_0}\right)-\frac{m_\pi}{2m_0}+\frac{m_\pi}{2m_0}\log\left(\frac{m_\pi}{m_0}\right)\right]\right\},
\end{equation}
\begin{equation}
R=\frac{1}{16\pi^2}\left\{\underline{-2+\log\left(\frac{m_0^2}{\mu^2}\right) }+\frac{m_\pi^2}{m_0^2}-\frac{2m_\pi}{m_0}\sqrt{1-\frac{m_\pi^2}{4 m_0^2}}\arccos\left(1-\frac{m_\pi^2}{2m_0^2}\right)\right\},
\end{equation}
where the conventional $\overline{\mathrm{MS}}$ procedure has been applied to remove the divergent pieces.
According to the chiral power-counting rule of Eq.~(\ref{eq:pcb}), the two underlined terms in the above equations break the power-counting and therefore must be removed. In the IR scheme, not only these two terms, but also
a series of higher-order terms are dropped. This can be clearly seen by expanding the regular part in $m_\pi$,
\begin{equation}
R=\frac{1}{16\pi^2}\left({-2+\log\left(\frac{m_0^2}{\mu^2}\right)}-\frac{m_\pi^2}{m_0^2}+\frac{m_\pi^4}{6 m_0^4}+\cdots\right).
\end{equation}
Furthermore, the infrared integral $I$ is not analytical at $m_\pi=2m_0$ because of the non-analyticity of the square root $\sqrt{4 m_0^2-m_\pi^2}$.

The infrared scheme has been reformulated by Schindler and collaborators~\cite{Schindler:2003xv} in a form analogous to the
EOMS scheme and has been applied to study the nucleon mass up to  $\mathcal{O}(p^6)$~\cite{Schindler:2006ha,Schindler:2007dr}, the axial, induced pseudoscalar, and pion-nucleon form factors up to $\mathcal{O}(p^4)$~\cite{Schindler:2006it}, and the
electromagnetic form factors of the nucleon up to $\mathcal{O}(p^4)$~\cite{Schindler:2005ke}.
To overcome the non-analyticity induced by the original IR scheme and to  establish direct
connections between the values of the LECs determined in the heavy-baryon ChPT and those obtained in the covariant BChPT,
Gail \textit{et al} has introduced the so-called modified infrared renormalization scheme, $\overline{\mathrm{IR}}$~\cite{Dorati:2007bk}. It has
recently been applied to study the finite-volume corrections to the electromagnetic current
 of the nucleon~\cite{Greil:2011aa} and the sigma terms and strangeness content of the octet baryons~\cite{Durr:2011mp}.

\subsection{The extended-on-mass-shell (EOMS) scheme}
The essential idea of the EOMS approach is that one can  calculate loop diagrams as usual in quantum field theory but then
drop the power-counting-breaking terms.  Physically, it is equivalent to a redefinition of the existing LECs. This procedure is always valid since by construction
ChPT contains all possible terms consistent with assumed symmetry principles~\cite{Weinberg:1978kz}. For instance, in Eq.~(\ref{eq:mass3}) the underlined PCB terms can be absorbed by
$m_0$ and $b_0+2b_F$. It should be noted that this is different from the infrared and heavy-baryon procedures, because they remove not only the PCB terms but also a series of higher-order
terms, which cannot be absorbed by the available LECs in question.

Technically, one has a few different ways to implement the EOMS scheme. One can first calculate a Feynman diagram, obtain the analytical results and then subtract the PCB terms. A second alternative is to expand
the integral in terms of some suitably defined small variables, perform the integration and obtain the PCB terms~\cite{Fuchs:2003qc}. One can also perform an expansion in terms of $1/m$ and identify the relevant PCB terms.  All three different implementations yield the same EOMS regularized result.
In the case of the integral Eq.~(\ref{eq:example}), one obtains
\begin{equation}
G_\mathrm{EOMS}=\frac{1}{16\pi^2}\left\{\frac{m_\pi^2}{m_0^2}\log\left(\frac{m_\pi}{m_0}\right)+\frac{2m_\pi}{m_0}\sqrt{1-\frac{m_\pi^2}{4m_0^2}}\arccos\left(\frac{m_\pi}{2m_0}\right)\right\},
\end{equation}
which has the same analytical structure as the original $\overline{\mathrm{MS}}$ result of Eq.(~\ref{eq:GMSbar}).

\subsection{A brief summary of the different formulations of BChPT}
\begin{figure}[t]
\begin{center}
\includegraphics[scale=0.16]{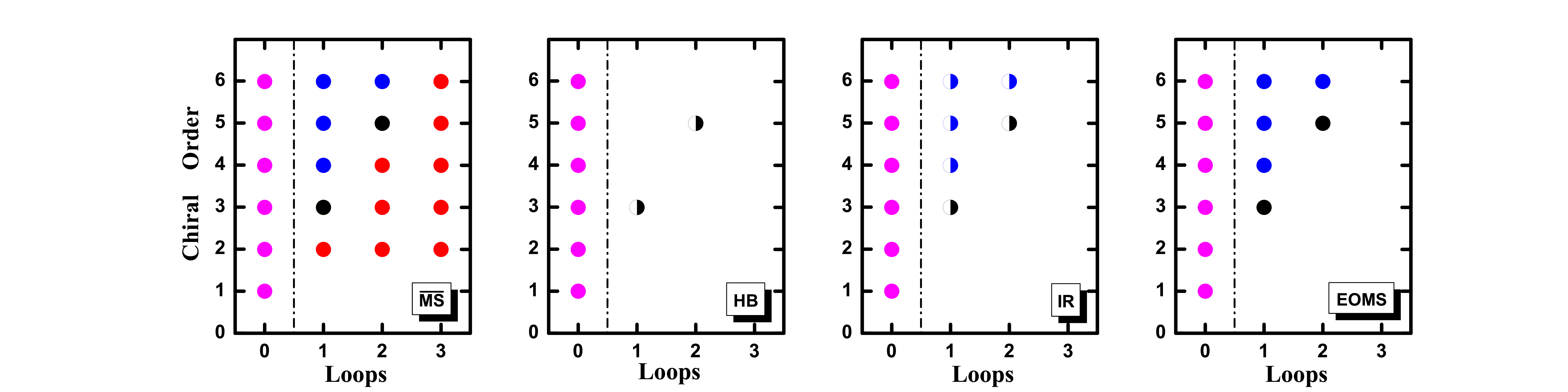}
\end{center}
\caption{(Color on-line) Chiral order vs. number of loops for pion-nucleon scattering in BChPT with $\overline{\mathrm{MS}}$, HB, IR, and EOMS
renormalization schemes.  Red filled circles denote PCB terms. Half-filled circles indicate the fact that the analytical terms of
the corresponding BChPT results  are not necessarily the same as those denoted by filled circles. \label{fig:renorm}}
\end{figure}

In Table \ref{table:4norm}, we summarize the main features of the four different formulations of BChPT.  All of them, except the $\overline{\mathrm{MS}}$ scheme,
satisfy the power-counting rule of Eq.~(\ref{eq:pcb}). The heavy-baryon ChPT contains only terms of the specified order, while both the IR and the
EOMS schemes contain a series of higher order terms, sometimes termed as recoil corrections. This is illustrated in Fig.~\ref{fig:renorm} for
the case of pion-nucleon scattering, where we have used filled and half-filled circles to denote the fact that at each chiral order the analytical
terms of different renormalization schemes for the same loop amplitude are not necessarily the same.

Furthermore, both the heavy-baryon and the infrared formulations spoil the analytical structure of loop amplitudes.
It was pointed out in Ref.~\cite{Bernard:1995dp} that in the infinite heavy baryon mass limit, the analytical structure of certain loop amplitudes (poles and cuts) may be disturbed. The same can be said about the infrared formulation. For instance, in the calculation of the nucleon mass, it can be explicitly shown that an unphysical cut at
$m_\pi=2 m_N$ is introduced~\cite{Pascalutsa:2011fp} (see also Section \ref{sec:ir}).

In addition, the separation of a covariant loop result into a singular part and a regular part may make both parts
divergent though the sum may be convergent. This is the case of the N$^3$LO result for the nucleon mass. In Table \ref{table:4norm}, we have used $(\surd)$
to denote this fact.
From Table \ref{table:4norm}, it can be concluded that formally the EOMS formulation is the most appealing one among the four schemes, $\overline{\mathrm{MS}}$, HB, IR, and EOMS. Furthermore, in practice, the EOMS formulation has recently been shown to converge relatively faster than the IR and the HB formulations. Some of these studies will be briefly reviewed in
the following section.


\begin{table*}
\caption{A comparison of different formulations of BChPT. \label{table:4norm}}
\begin{ruledtabular}
\begin{tabular}{ccccc}
\footnotesize
 & power-counting & covariance & analyticity & ultraviolet regularisation \\
\hline
$\overline{\mathrm{MS}}$ & $-$ & $\surd$ & $\surd$ & $\surd$ \\
HB & $\surd$ & $-$ & $-$ & $\surd$ \\
IR &  $\surd$ &  $\surd$  &  $-$ & $(\surd)$\\
EOMS  & $\surd$ & $\surd$ & $\surd$  & $\surd$  \\
\end{tabular}
\end{ruledtabular}
\end{table*}

\section{Recent applications in the $u$, $d$, and $s$ three-flavor sector}

In this section, we briefly review some recent applications of the covariant (EOMS) BChPT in the one-baryon sector.~\footnote{It should be noted that the EOMS formulation has  been applied to the nucleon-nucleon system~\cite{Epelbaum:2012ua}.} In particular, we focus on
the three-flavor sector of $u$, $d$, and $s$ quarks. In this short review, we limit our discussions to the
  EOMS BChPT. It should be pointed out that despite of having the analyticity problem and being slow in convergence even compared to
  the heavy-baryon ChPT,
  the infrared BChPT has been applied to study a variety of physical observables in the three-flavor sector, e.g.,
the ground-state octet baryon masses up to $\mathcal{O}(p^3)$~\cite{Ellis:1999jt} and $\mathcal{O}(p^4)$ ~\cite{Frink:2004ic},
   the baryon axial currents  up to $\mathcal{O}(p^3)$~\cite{Zhu:2000zf},
   the baryon electromagnetic form factors up to $\mathcal{O}(p^4)$~\cite{Kubis:2000aa}, 
     the hyperon decay form factors up to $\mathcal{O}(p^4)$~\cite{Lacour:2007wm}, and the meson-baryon scattering lengths up to leading one-loop order~\cite{Mai:2009ce}. Recently, the 
    first PDF moments~\cite{Bruns:2011sh}  and the electric dipole moments~\cite{Guo:2012vf} of the ground-state octet baryons were also calculated up to leading one-loop order. 
    All these studies have not explicitly considered the contributions of the virtual decuplet baryons. Furthermore, these studies yielded controversial results regarding the
    convergence behaviour of the infrared BChPT. For instance, the convergence behaviour of the axial couplings was found to be problematic in Ref.~\cite{Zhu:2000zf}, but
 the convergence behaviour of the hyperon charge radii was found to be more than satisfactory in Ref.~\cite{Kubis:2000aa}. 
\subsection{Magnetic moments of the octet and decuplet baryons}

\subsubsection{Virtual octet contributions to the magnetic moments of octet baryons}

In the SU(3) flavor symmetric limit one can relate the magnetic moments of the baryon-octet and the $\Lambda\Sigma^0$ transition to those of the proton and the neutron. These are the celebrated Coleman-Glashow formulas~\cite{Coleman:1961jn}. To properly implement SU(3) breaking, ChPT should be an appropriate framework to tackle this problem in a systematic fashion. This was first attempted by Caldi and Pages even before ChPT as we know today was formulated~\cite{Caldi:1974ta} and in the HBChPT framework by Jenkins et al.~\cite{Jenkins:1992pi}.  It was found that the leading order  SU(3) breaking effects induced by loops are too large and worsen the SU(3) symmetric descriptions. This problem has often been used to question the validity of SU(3) ChPT in the one-baryon sector~\cite{Durand:1997ya,Meissner:1997hn,Donoghue:2004vk}.

In the last decade several calculations in the HBChPT  up to next-to-next-to-leading order (NNLO) have been performed both with~\cite{Jenkins:1992pi,Durand:1997ya,Puglia:1999th} and without~\cite{Meissner:1997hn} the inclusion of the baryon decuplet. The large number of LECs appearing at this order reduces the predictive power of the theory.
The baryon-octet magnetic moments have been calculated using the IR method~\cite{Kubis:2000aa} and, at NLO, the SU(3)-breaking corrections are still large. Moreover, the agreement with the data is even worse than in the HBChPT. The sizes of the NLO terms raise the question about the convergence of the chiral series~\cite{Durand:1997ya,Meissner:1997hn,Donoghue:2004vk}.

\begin{table*}
\caption{Magnetic moments of the octet baryons
calculated in leading-order BChPT, next-to-leading order BChPT formulated in the
HB, IR and EOMS schemes.  All the values for the magnetic moments are expressed in units of nuclear magnetons, while $\tilde{b}_6^D$ and $\tilde{b}_6^F$ are dimensionless. Taken from
Ref.~\cite{Geng:2008mf}. \label{table:mag}}
\begin{ruledtabular}
\begin{tabular}{ccccccccccccc}
\footnotesize
 & $p$ & $n$ & $\Lambda$ & $\Sigma^-$ & $\Sigma^+$ & $\Sigma^0$ & $\Xi^-$ & $\Xi^0$ & $\Lambda\Sigma^0$ & $\tilde{b}_6^D$ & $\tilde{b}_6^F$ & $\tilde{\chi}^2$ \\
\hline
\multicolumn{13}{c}{$\mathcal{O}(p^2)$}  \\
\hline
Tree level & 2.56 & -1.60 & -0.80 & -0.97 & 2.56 & 0.80 & -1.60 & -0.97 & 1.38 & 2.40 & 0.77 & 0.46  \\
\hline
\multicolumn{13}{c}{$\mathcal{O}(p^3)$}  \\
\hline
HB  & 3.01 & -2.62 & -0.42 & -1.35 & 2.18 & 0.42 & -0.70 & -0.52 & 1.68 & 4.71 & 2.48 & 1.01  \\

IR  & 2.08 & -2.74 & -0.64 & -1.13 & 2.41 & 0.64 & -1.17 & -1.45 & 1.89 & 4.81 & 0.012 & 1.86 \\

EOMS & 2.58 & -2.10 & -0.66 & -1.10 & 2.43 & 0.66 & -0.95 & -1.27 & 1.58 & 3.82 & 1.20 & 0.18  \\
\hline
Exp. &  2.793(0) & -1.913(0) & -0.613(4) & -1.160(25) & 2.458(10) & --- & -0.651(3) &-1.250(14) & $\pm$  1.61(8) & \multicolumn{3}{c}{---}  \\
\end{tabular}
\end{ruledtabular}
\end{table*}

A study of the magnetic moments of octet baryons in the EOMS formulation was performed in Refs.~\cite{Geng:2008mf,Geng:2009hh}.
Up to NLO, there are two LECs appearing in the BChPT calculation, i.e., $\tilde{b}_6^D$ and $\tilde{b}_6^F$. One can  fix them
by performing a fit of the BChPT results to the corresponding experimental data. The results are shown in
Table \ref{table:mag}, where $\tilde{\chi}^2$ is defined as $\tilde{\chi}^2=\sum (\mu_{th}-\mu_{exp})^2$. One should note that
this is not a proper definition of $\chi^2$, but is used nevertheless to compare with previous studies. Clearly, unlike HB and IR, the EOMS results improve the
tree level results. On the other hand, the IR results are even more off the data than the HB results.

Furthermore, in Ref.~\cite{Geng:2008mf}, it was pointed out that the EOMS results show a better convergence behaviour. To see this, one can separate the $\mathcal{O}(p^2)$ contribution
from that of $\mathcal{O}(p^3)$:
\begin{equation}
\begin{array}{ll}
\mu_p=3.47\,\left( 1-0.257\right),\quad& \mu_n=-2.55\, \left( 1-0.175\right), \nonumber\\
\mu_\Lambda=-1.27\, \left(1-0.482\right),\quad& \mu_{\Sigma^-}=-0.93\,\left( 1+0.187 \right), \nonumber\\
\mu_{\Sigma^+}=3.47\,\left( 1-0.300 \right),\quad& \mu_{\Sigma^0}=1.27\, \left(1-0.482\right),\nonumber\\
\mu_{\Xi^-}=-0.93\,\left( 1+0.025 \right), \quad&\mu_{\Xi^0}=-2.55\,\left( 1-0.501 \right),\nonumber\\
\mu_{\Lambda\Sigma^0}=2.21\,\left( 1- 0.284\right),\nonumber
\end{array}
\end{equation}
where the first number in the parenthesis indicates the LO contribution and the second number denotes the percentage of the NLO contribution relative to that of the LO.
It is clear that in the EOMS framework the NLO contribution is at most about 50\% of the LO, while the same number can be up to $70\%$ in HB or $300\%$ in IR.

In the chiral limit, all three approaches should give the same description as the tree level results, while in the real world only the EOMS results improve the tree level description. Such an evolution is shown
in Fig.~2.
\begin{figure}[t]
\includegraphics[scale=0.65]{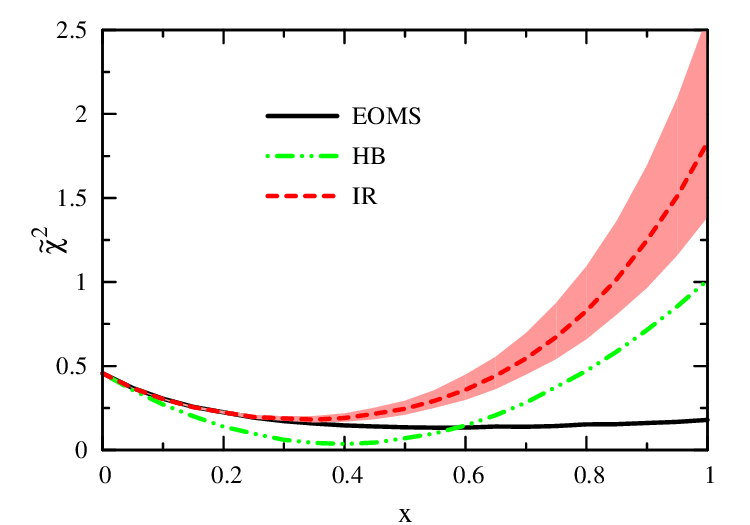}
\caption{(Color on-line) SU(3)-breaking evolution of the minimal $\tilde{\chi}^2$ in the $\mathcal{O}(p^3)$ BChPT approaches.  The shaded areas  are
produced by  varying $M_B$ from 0.8 GeV to 1.1 GeV. This effect lies within the line thickness in the EOMS case, while the HB is insensitive to it.   Taken from Ref.~\cite{Geng:2008mf}. \label{fig_graph}}
\end{figure}

\subsubsection{Virtual decuplet contributions to the magnetic moments of the octet baryons}
A basic assumption of EFTs is that high-energy dynamics can be integrated out or, in other words, their effects can be approximated by contact interactions. In the present case,
this means that in the loops virtual baryons except the octet baryons  can be neglected. In the three-flavor sector, the above assumption may not be valid.
The reason is that the lowest-lying decuplet is just, on average, 231 MeV above the ground-state octet. This energy difference is slightly larger than the pion mass and much smaller than the kaon mass.  Therefore one needs to check the contributions of the virtual decuplet baryons. Such a work
was performed in Ref.~\cite{Geng:2009hh}.

 Inclusion of the decuplet baryons is not as straightforward as one would  like because
the description of higher-spin ($s\geq$3/2) particles in a relativistic quantum field theory is known to be problematic because of the presence of unphysical lower-spin components. For instance, in the Rarita-Schwinger (RS) formulation~\cite{Rarita:1941mf} adopted in this work, the field representation of a massive 3/2-particle is a vector-spinor $\psi_\mu$ with two unphysical spin-1/2 components in addition to the spin-3/2 components. In the presence of interactions the unphysical degrees of freedom are known to lead to pathologies like non-positive definite commutators or acausal propagation for the coupling of the photon~\cite{Johnson:1960vt,Velo:1969bt,Deser:2000dz}. Equivalent problems in phenomenological hadronic interactions have also been extensively discussed~\cite{Nath:1971wp,Hagen:1972ea,Singh:1973gq,Pascalutsa:1998pw,Pascalutsa:1999zz}. In the context of ChPT one can use field redefinitions on the conventional chiral Lagrangians in order to cast the interactions in a form that is invariant under the transformation $\psi_\mu\rightarrow\psi_\mu+\partial_\mu\epsilon$~\cite{Pascalutsa:2000kd,Pascalutsa:2006up,Krebs:2008zb}. The resulting gauge symmetry ensures to keep active only the physical degrees of freedom~\cite{Pascalutsa:1998pw}.
Furthermore, there is abundant work concerning the inclusion of spin-3/2 resonances in the framework of baryonic effective field theories~\cite{Tang:1996sq,Hemmert:1997ye,Hacker:2005fh,Wies:2006rv}.

The baryon-decuplet consists of a SU(3)-flavor multiplet of spin-3/2 resonances that can be represented with the Rarita-Schwinger field $T_\mu\equiv T^{ade}_\mu$ with the following associations:
$T^{111}=\Delta^{++}$, $T^{112}=\Delta^+/\sqrt{3}$,
$T^{122}=\Delta^0/\sqrt{3}$, $T^{222}=\Delta^-$, $T^{113}=\Sigma^{*+}/\sqrt{3}$,
$T^{123}=\Sigma^{*0}/\sqrt{6}$, $T^{223}=\Sigma^{*-}/\sqrt{3}$,
$T^{133}=\Xi^{*0}/\sqrt{3}$, $T^{233}=\Xi^{*-}/\sqrt{3}$, and $T^{333}=\Omega^-$. The covariantized free Lagrangian is
\begin{equation}
 \mathcal{L}_{D}=\bar{T}^{abc}_\mu(i\gamma^{\mu\nu\alpha}D_\alpha-M_D\gamma^{\mu\nu})T^{abc}_\nu, \label{Eq:RSLag}
\end{equation}
with $M_D$ the chiral limit decuplet-baryon mass and $D_\nu T_\mu^{abc}=\partial_\nu T_\mu^{abc}+(\Gamma_\nu)_d^a T_\mu^{dbc}
+(\Gamma_\nu)_d^b T_\mu^{adc}+(\Gamma_\nu)_d^c T_\mu^{abd}$. In the last and following Lagrangians we sum over any repeated SU(3)-index denoted by latin characters $a,b,c,\ldots$, and $(X)^a_b$ denotes the element of row $a$ and column $b$ of the matrix representation of $X$.

The conventional lowest-order chiral Lagrangian for the interaction of the decuplet- and octet-baryons with the pseudoscalar mesons expanded up to one meson field is
\begin{eqnarray}
 \mathcal{L}^{(1)}_{\phi BD}=\frac{\mathcal{C}}{F_\phi}\;\varepsilon^{abc}\bar{T}^{ade}_\mu\left(g^{\mu\nu}+z \gamma^\mu\gamma^\nu\right)
 B^e_c\,\partial_\nu\phi^d_b+{\rm h.c.},\label{Eq:CnvLag}
\end{eqnarray}
where $\mathcal{C}$ is the pseudoscalar meson-octet baryon-decuplet baryon ($\phi B D$) coupling, $F_\phi$ the meson-decay constant and $z$ is an off-shell parameter. An analysis of the constraint structure of the interacting theory of Eqs. (\ref{Eq:RSLag}, \ref{Eq:CnvLag}) yields $z=-1$~\cite{Nath:1971wp}.~\footnote{See, however, Ref.~\cite{Benmerrouche:1989uc} for a dispute against such a choice.} Nevertheless, the resulting interaction leads to well-known problems afflicting the relativistic quantum field theory of 3/2-spinors~\cite{Hagen:1972ea,Singh:1973gq,Pascalutsa:1998pw}.
\begin{table*}
\centering
\caption{Baryon octet magnetic moments in BChPT up to $\mathcal{O}(p^3)$.  The SU(3)-symmetric description is compared with the different $\mathcal{O}(p^3)$ BChPT calculations, i.e.,  the HB and the EOMS results both with (O+D) and without (O) the inclusion of dynamical decuplet baryons. In the covariant case the numerical results are obtained using the consistent couplings (\ref{Eq:CnsLag}) and the conventional couplings (\ref{Eq:CnvLag}) with $z=-1$. The experimental values are from Ref.~\cite{Amsler:2008zzb}.
Taken from Ref.~\cite{Geng:2009hh}. \label{table:Results}}
\begin{tabular}{c|c|cc|ccc|c|}
\cline{2-8}
& &\multicolumn{2}{|c|}{Heavy Baryon $\mathcal{O}(p^3)$}&\multicolumn{3}{|c|}{Covariant EOMS $\mathcal{O}(p^3)$}& \\
\cline{3-7}
&  \raisebox{1ex}[0pt]{Tree level $\mathcal{O}(p^2)$}& O & O+D& O & O+D (conv.) &O+D (consist.) & \raisebox{1ex}[0pt]{Expt.}\\
\hline\hline
\multicolumn{1}{|c|}{\textit{p}} & 2.56 &3.01& 3.47 & 2.60 & 3.18 & 2.61& 2.793(0)\\
\multicolumn{1}{|c|}{\textit{n}} & -1.60 &-2.62&-2.84& -2.16 & -2.51&-2.23 &-1.913(0)\\
\multicolumn{1}{|c|}{$\Lambda$}& -0.80 &-0.42&-0.17&-0.64&-0.29 &-0.60&-0.613(4) \\
\multicolumn{1}{|c|}{$\Sigma^-$}& -0.97 &-1.35&-1.42& -1.12 &-1.26&-1.17&-1.160(25)\\
\multicolumn{1}{|c|}{$\Sigma^+$}& 2.56 &2.18&1.77& 2.41& 1.84 &2.37&2.458(10) \\
\multicolumn{1}{|c|}{$\Sigma^0$}& 0.80 &0.42&0.17& 0.64 & 0.29& 0.60 & ... \\
\multicolumn{1}{|c|}{$\Xi^-$}& -1.60 &-0.70&-0.41& -0.93 & -0.78& -0.92 & -0.651(3) \\
\multicolumn{1}{|c|}{$\Xi^0$}&-0.97 &-0.52&-0.56& -1.23 & -1.05& -1.22 & -1.250(14) \\
\multicolumn{1}{|c|}{$\Lambda\Sigma^0$} & 1.38 &1.68&1.86& 1.58 & 1.88 & 1.65& $\pm$1.61(8)\\
\hline
\multicolumn{1}{|c|}{$b_6^D$}& 2.40&4.71&5.88 & 3.92 & 5.76 & 4.30 & \\
\multicolumn{1}{|c|}{$b_6^F$}& 0.77 &2.48&2.49& 1.28 & 1.03 & 1.03 & ...\\
\multicolumn{1}{|c|}{$\bar{\chi}^2$}&0.46&1.01&2.58& 0.18 & 1.06 & 0.22& \\
\hline
\end{tabular}
\end{table*}

The alternative approach of demanding the effective Lagrangians to be spin-3/2-gauge invariant leads, after a field redefinition, to the ``consistent'' $\phi B D$ interaction ~\cite{Pascalutsa:1998pw,Pascalutsa:1999zz}
\begin{equation}
\mathcal{L}\,'^{\,(1)}_{\phi B D}=\frac{i\,\mathcal{C}}{M_D F_\phi}\;\varepsilon^{abc}\left(\partial_\alpha\bar{T}^{ade}_\mu\right)\gamma^{\alpha\mu\nu}
 B^e_c\,\partial_\nu\phi^d_b+{\rm h.c.},\label{Eq:CnsLag}
\end{equation}
which is on-shell equivalent to Eq. (\ref{Eq:CnvLag}).
In addition, one obtains a second-order $\phi\phi B B$ contact term
\begin{equation}
\mathcal{L}^{(2)}_{\phi\phi B B}=\frac{\mathcal{C}^2}{12M_D^2 F_\phi^2}\left(3\langle \bar{B}\{[\partial_\mu\phi,\partial_\nu\phi],(R^{\mu\nu}B)\}\rangle+\langle \bar{B}[[\partial_\mu\phi,\partial_\nu\phi],(R^{\mu\nu}B)]\rangle-6\langle\bar{B}\partial_\mu\phi\rangle\langle\partial_\nu\phi (R^{\mu\nu}B)\rangle\right),\label{Eq:HOCTLag}
\end{equation}
where $R^{\mu\nu}=i\gamma^{\mu\nu\alpha}\partial_\alpha+M_D\gamma^{\mu\nu}$ and $\langle\ldots\rangle$ denotes the trace in flavor space. The latter Lagrangian is interpreted as carrying the spin-1/2 content of the Lagrangian (\ref{Eq:CnvLag}). This term is eliminated by absorbing it into suitable higher-order LECs (for a relevant discussion, see also Refs.~\cite{Tang:1996sq,Krebs:2009bf}).

Using the EOMS scheme to remove the power-counting-breaking terms in the loops and the so-called small scale expansion~\cite{Hemmert:1997ye} to treat
the octet and decuplet mass difference, one obtains the results shown in Table \ref{table:Results}. Immediately, one notices that in the ``consistent'' decuplet-octet coupling scheme, the inclusion of the virtual decuplet baryons only slightly worsens
the octet-only description. On the other hand, using the conventional coupling scheme of the decuplet-octet couplings, the inclusion of the virtual decuplet
deteriorates a lot the description of the magnetic moments of the octet baryons.

\begin{figure}[h]
\includegraphics[width=\columnwidth]{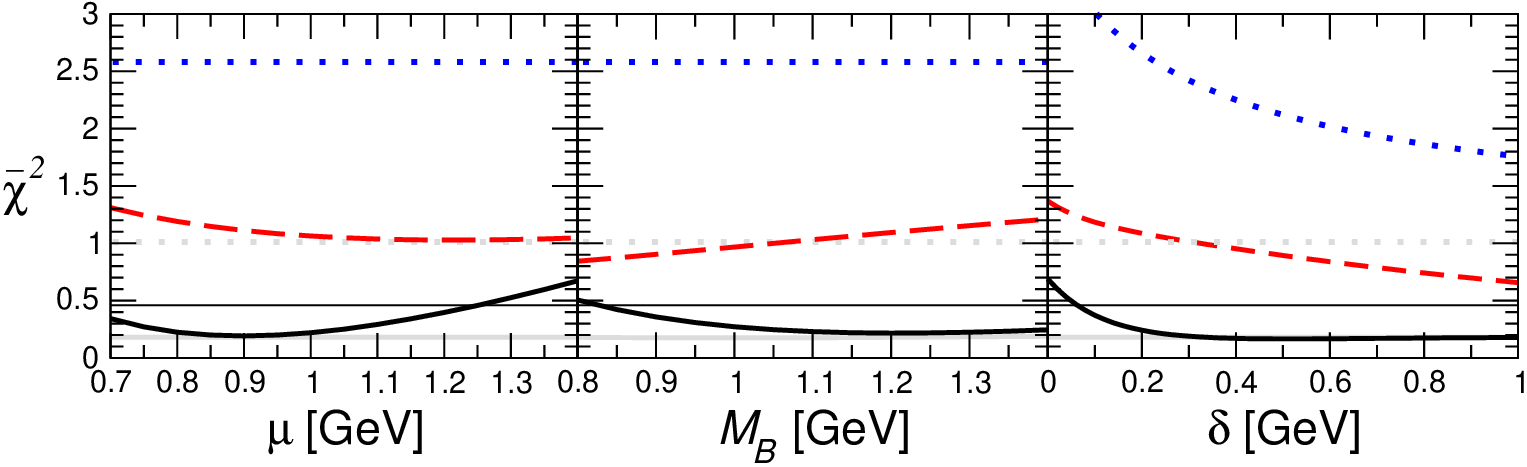}
\caption{Uncertainties of the numerical results of Table \ref{table:Results} due to the values of the regularisation scale $\mu$ (\textit{left} panel), the average baryon mass $M_B$ (\textit{center} panel) and the decuplet-octet mass splitting $\delta$ (\textit{right} panel). The lines represent the $\bar{\chi}^2$ for the results obtained in HB (dotted line) and in the EOMS approach using conventional (dashed line) and consistent (thick solid line) couplings. The grey lines represent the $\bar{\chi}^2$ for the case without explicit decuplet resonances in HB (dotted) and in the EOMS formulation (solid). For reference  the SU(3)-symmetric description (thin solid line) is also shown.  Taken from Ref.~\cite{Geng:2009hh} \label{fig:smag}}
\end{figure}
It has been checked that the conclusion does not depend sensitively on the values of the regularisation scale, the decuplet-octet mass splitting, and the average baryon mass, as shown in
Fig.~\ref{fig:smag}.

\subsubsection{Magnetic moments of the decuplet baryons}
The EOMS BChPT has also been applied to study the electromagnetic structure of the decuplet baryons~\cite{Geng:2009ys}. The structure of the spin-3/2 particles, as probed by photons, is encoded into four electromagnetic form factors~\cite{Nozawa:1990gt}:
\begin{equation}
\langle T(p')|J^\mu|T(p)\rangle=-\bar{u}_\alpha(p')\Big\{\Big[F_1^*(\tau)\gamma^\mu+\frac{i\sigma^{\mu\nu}q_\nu}{2M_D}F_2^*(\tau)\Big]g^{\alpha\beta}+\Big[F_3^*(\tau)\gamma^\mu+\frac{i\sigma^{\mu\nu}q_\nu}{2M_D}F_4^*(\tau)\Big]\frac{q^\alpha q^\beta}{4M_D^2}\Big\}u_\beta(p),  \label{Eq:DeltaFF0}
\end{equation}
where $u_\alpha$ are the Rarita-Schwinger spinors and $\tau=-q^2/(4M_D^2)$. One can define the electric monopole and quadrupole, and the magnetic dipole and octupole form factors in terms of the $F_i^*$'s:
\begin{eqnarray}
&&G_{E0}(\tau)=(F_1^*(\tau)-\tau F_2^*(\tau))+\frac{2}{3}\tau G_{E2}(\tau),\label{Eq:GEO}\\
&&G_{E2}(\tau)=(F_1^*(\tau)-\tau F_2^*(\tau))-\frac{1}{2}(1+\tau)(F_3^*(\tau)-\tau F_4^*(\tau)),\label{Eq:GE2}\\
&&G_{M1}(\tau)=(F_1^*(\tau)+F_2^*(\tau))+\frac{4}{5}\tau G_{M3}(\tau), \label{Eq:GM1}\\
&&G_{M3}(\tau)=(F_1^*(\tau)+F_2^*(\tau))-\frac{1}{2}(1+\tau)(F_3^*(\tau)+F_4^*(\tau)).  \label{Eq:GM3}
\end{eqnarray}
At $q^2=0$, the multipole form factors define the static electromagnetic moments, namely, the charge $Q$, the magnetic dipole moment $\mu$, the electric quadrupole moment $\mathcal{Q}$, and the magnetic octupole moment $O$
\begin{eqnarray}
&&Q=G_{E0}(0)=F_1^*(0),  \label{Eq:Dcharge}\\
&&\mu=\frac{e}{2M_D}G_{M1}(0)= \frac{e}{2M_D}(Q+F_2^*(0)), \label{Eq:DMDM}\\
&&\mathcal{Q}=\frac{e}{M_D^2}G_{E2}(0)=\frac{e}{M_D^2}(Q-\frac{1}{2}F_3^*(0)),\label{Eq:DQEM}\\
&&O=\frac{e}{2M_D^3}G_{M3}(0)=\frac{e}{2M_D^3}\left(G_{M1}(0)-\frac{1}{2}(F_3^*(0)+F_4^*(0))\right).\label{Eq:DOMM}
\end{eqnarray}

The electromagnetic multipole moments of the spin-3/2 resonances are connected with their spatial electromagnetic distributions and, therefore, with their internal structure. Particularly, the electric quadrupole moment and magnetic octupole moment measure the departure from a spherical shape of the charge and from a dipole magnetic distribution, respectively.

Besides the static electromagnetic moments, the slope of the form factors at $q^2=0$ is also of phenomenological interest. In particular the one corresponding to $G_{E0}$ is the so-called squared CR:
\begin{equation}
\langle r_{E0}^2\rangle=6\frac{dG_{E0}(q^2)}{dq^2}\Big|_{q^2=0}=6\frac{dF^*_{1}(q^2)}{dq^2}\Big|_{q^2=0}+\frac{3}{2M_D^2}F^*_2(0)-\frac{1}{M_D^2}G_{E2}(0).\label{Eq:ChRadiusDef}
\end{equation}

\begin{table*}[t]
\centering
\caption{Magnetic dipole moments of the decuplet baryons (in nuclear magnetons)  in $\mathcal{O}(p^3)$ EOMS BChPT, in comparison with the SU(3)-symmetric description and with those obtained in other theoretical approaches including the NQM~\cite{Hikasa:1992je}, the RQM~\cite{Schlumpf:1993rm}, the $\chi$QM ~\cite{Wagner:2000ii}, the $\chi$QSM~\cite{Ledwig:2008es}, the QCD-SR~\cite{Lee:1997jk}, (extrapolated) lQCD~\cite{Leinweber:1992hy,Lee:2005ds}, large $N_c$~\cite{Luty:1994ub} and the HBChPT calculation of Ref.~\cite{Butler:1993ej}. The experimental values are also included for reference~\cite{Yao:2006px}.  Taken from Ref.~\cite{Geng:2009ys}. \label{Table:ResMDM}}
\begin{ruledtabular}
\begin{tabular}{ccccccccccc}
\footnotesize
&$\Delta^{++}$&$\Delta^+$&$\Delta^0$&$\Delta^-$&$\Sigma^{*+}$&$\Sigma^{*0}$&$\Sigma^{*-}$&$\Xi^{*0}$&$\Xi^{*-}$&$\Omega^-$\\
\hline
\multicolumn{1}{c}{SU(3)-symm.}&4.04&2.02&0&-2.02&2.02&0&-2.02&0&-2.02&-2.02\\
\multicolumn{1}{c}{NQM~\cite{Hikasa:1992je}}&5.56&2.73&-0.09&-2.92&3.09&0.27&-2.56&0.63&-2.2&-1.84\\
\multicolumn{1}{c}{RQM~\cite{Schlumpf:1993rm}}&4.76&2.38&0&-2.38&1.82&-0.27&-2.36&-0.60&-2.41&-2.35\\
\multicolumn{1}{c}{$\chi$QM~\cite{Wagner:2000ii}}&6.93&3.47&0&-3.47&4.12&0.53&-3.06&1.10&-2.61&-2.13\\
\multicolumn{1}{c}{$\chi$QSM~\cite{Ledwig:2008es}} &4.85&2.35&-0.14&-2.63&2.47&-0.02&-2.52&0.09&-2.40&-2.29\\
\multicolumn{1}{c}{QCD-SR~\cite{Lee:1997jk}}&4.1(1.3)&2.07(65)&0&-2.07(65)&2.13(82)&-0.32(15)&-1.66(73)&-0.69(29)&-1.51(52)&-1.49(45)\\
\multicolumn{1}{c}{lQCD~\cite{Leinweber:1992hy}}&6.09(88)&3.05(44)&0&-3.05(44)&3.16(40)&0.329(67)&-2.50(29)&0.58(10)&-2.08(24)&-1.73(22)\\
\multicolumn{1}{c}{lQCD~\cite{Lee:2005ds}}&5.24(18)&0.97(8)&-0.035(2)&-2.98(19)&1.27(6)&0.33(5)&-1.88(4)&0.16(4)&-0.62(1)&---\\
\multicolumn{1}{c}{large $N_c$~\cite{Luty:1994ub}}&5.9(4)&2.9(2)&---&-2.9(2)&3.3(2)&0.3(1)&-2.8(3)&0.65(20)&-2.30(15)&-1.94\\
\multicolumn{1}{c}{HBChPT~\cite{Butler:1993ej}}&4.0(4)&2.1(2)&-0.17(4)&-2.25(19)&2.0(2)&-0.07(2)&-2.2(2)&0.10(4)&-2.0(2)&-1.94\\
\hline
\multicolumn{1}{c}{EOMS BChPT}&6.04(13)&2.84(2)&-0.36(9)&-3.56(20)&3.07(12)&0&-3.07(12)&0.36(9)&-2.56(6)&-2.02\\
\multicolumn{1}{c}{Expt.~\cite{Yao:2006px}}&5.6$\pm$1.9&$2.7^{+1.0}_{-1.3}\pm1.5\pm3$&---&---&---&---&---&---&---&-2.02$\pm0.05$\\
\end{tabular}
\end{ruledtabular}
\end{table*}
At next-to-leading order, there is only unknown LEC in BChPT, $g_d$, for the magnetic dipole moment, which can be fixed by reproducing the magnetic moment of the $\Omega^-$. Once this is done, BChPT can make parameter-free predictions
for the magnetic moments of all the other decuplet baryons. The results are shown in Table \ref{Table:ResMDM} and compared with those of a number of other studies, including
the non-relativistic quark model (NQM)~\cite{Hikasa:1992je}, the relativistic quark model (RQM)~\cite{Schlumpf:1993rm}, the chiral quark model ~\cite{Wagner:2000ii},
the chiral quark soliton model~\cite{Ledwig:2008es}, the QCD sum rule approach~\cite{Lee:1997jk}, the LQCD approach~\cite{Leinweber:1992hy,Lee:2005ds}, the Large $N_c$ approach~\cite{Luty:1994ub}, the
NLO heavy-baryon ChPT~\cite{Butler:1993ej}. It is clear that the EOMS BChPT results are in very good agreement with the data.

Up to next-to-leading order, there is only one unknown LEC $g_q$ for the electric quadrupole moments, and
one unknown LEC $g_{\mathrm{cr}}$ for the charge radii of the decuplet baryons.~\footnote{The magnetic
octupole moments depend on $g_q$ and $g_d$.}  If the electric quadrupole moment and
the charge radius of a decuplet baryon is known via either experiment or LQCD, the unknown LECs can be fixed and
one can make predictions for all the other decuplet baryons. The results then serve as
a stringent test on the validity of covariant BChPT at next-to-leading order. This has been explored in Ref.~\cite{Geng:2009ys}.
\subsection{Masses and sigma terms of the octet baryons}
\subsubsection{Masses of the octet baryons in N$^3$LO EOMS BChPT}
Recently, the lowest-lying baryon spectrum, composed of up, down and strange quarks, has been studied by various LQCD collaborations~\cite{Alexandrou:2009qu,Durr:2008zz,Aoki:2008sm,Aoki:2009ix,WalkerLoud:2008bp, Lin:2008pr,Bietenholz:2010jr,Bietenholz:2011qq,Beane:2011pc}.  Such studies not only improve our understanding of some fundamental questions in physics, such as the origin of mass~\cite{Wilczek:2012sb},
but also provide a benchmark for future studies based on lattice QCD.  In addition, such quark mass, volume, lattice spacing dependent simulations also enable us to better understand
different aspects of QCD, which in reality has fixed quark mass and is continuous and in infinite space time.  The latter perspective also allows one to fix the many unknown LECs of BChPT, which can not be easily determined, even if possible, by experimental data alone.

However, because these calculations adopt different lattice setup and all of them lead to the same continuum theory, it is crucial to test whether the results for baryon masses are consistent with each other~\cite{Beringer:1900zz}. On the other hand, since lattice QCD simulations are performed in a finite hypercube and  with larger than physical light-quark masses~\cite{Fodor:2012gf}, the final results can only be obtained by extrapolating to the physical point (chiral extrapolation) and infinite space-time (finite volume corrections). ChPT provides a useful framework to perform such extrapolations and to study the induced uncertainties.

In the past decades, the ground-state (g.s.) octet baryon masses have been studied extensively~\cite{Jenkins:1991ts,Bernard:1993nj,Banerjee:1994bk,Borasoy:1996bx,WalkerLoud:2004hf, Ellis:1999jt,Frink:2004ic,Frink:2005ru,Lehnhart:2004vi,MartinCamalich:2010fp,Young:2009zb, Semke:2005sn,Semke:2007zz,Semke:2011ez,Bruns:2012eh,Lutz:2012mq}.
It is found that SU(3) heavy-baryon ChPT converges rather slowly~\cite{WalkerLoud:2008bp,Ishikawa:2009vc}\footnote{It was shown in Refs.~\cite{Jenkins:2009wv,WalkerLoud:2011ab}
that the various linear combinations of
baryon masses, chosen to have definite scaling in terms of $1/N_c$ and SU(3) symmetry breaking,
can be  better described by the combination of the large $N_c$  and SU(3) expansions.}. Furthermore, most calculations are performed only up to NNLO because of the many unknown LECs at N$^3$LO except those of Refs.~\cite{Borasoy:1996bx,WalkerLoud:2004hf,Frink:2004ic,Semke:2011ez,Bruns:2012eh,Lutz:2012mq}. Regarding chiral extrapolations, Young and Thomas~\cite{Young:2009zb} obtained very good
results using the FRR scheme up to NNLO by fitting the LHPC \cite{WalkerLoud:2008bp} and PACS-CS \cite{Aoki:2008sm} lattice data. In Ref.~\cite{MartinCamalich:2010fp}, we applied the NNLO EOMS BChPT to analyze the same lattice data and found that the EOMS BChPT can provide a better description of lattice data and is more suitable for chiral extrapolation purposes than heavy-baryon ChPT and next-to-leading order BChPT. Recently, using a partial summation scheme up to N$^3$LO, Semke and Lutz~\cite{Semke:2011ez,Semke:2012gs,Lutz:2012mq} found that the BMW~\cite{Durr:2008zz}, HSC~\cite{Lin:2008pr}, PACS-CS~\cite{Aoki:2008sm}, LHPC~\cite{WalkerLoud:2008bp}, and QCDSF-UKQCD~\cite{Bietenholz:2011qq} lattice results can be well described.  It should be noted, however, that Semke and Lutz adopted the large $N_c$ argument to fix the values of some LECs and neglected the SU(3)-symmetric contributions.

On the other hand, until very recently, a simultaneous description of all the $n_f=2+1$ lattice data with finite-volume effects taken into account self-consistently is still missing. Such a study is necessary for a clarification of the convergence problem and for testing the consistency between different lattice simulations. Furthermore, it also provides a good opportunity to determine/constrain the many unknown LECs of BChPT at N$^3$LO.

\begin{figure}[b]
\centering
\includegraphics[width=15cm]{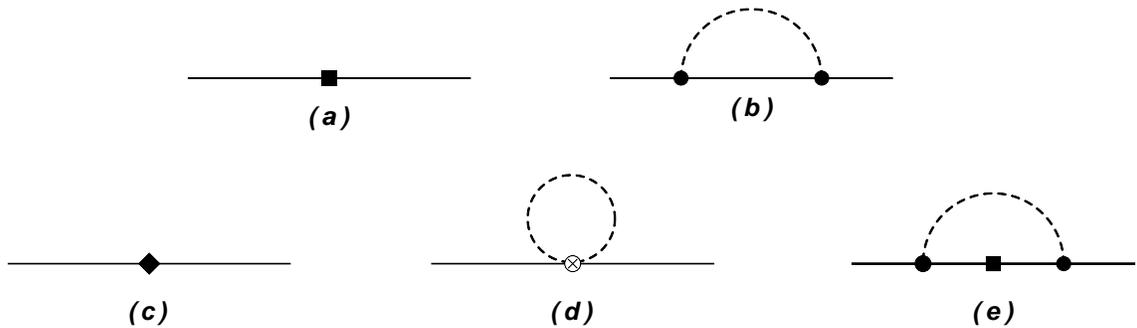}
\caption{Feynman diagrams contributing to the octet-baryon masses up to $\mathcal{O}(p^4)$ in the EOMS BChPT. The solid lines correspond to octet-baryons and dashed lines refer to NGBs. The black boxes (diamonds) indicate second (fourth) order couplings. The solid dot (circle-cross) indicates an insertion from the dimension one (two) meson-baryon Lagrangians. Wave function renormalization diagrams are not explicitly shown but included in the calculation.}
\label{dia:mass4}
\end{figure}
\begin{figure}[t]
\centering
\includegraphics[width=15cm]{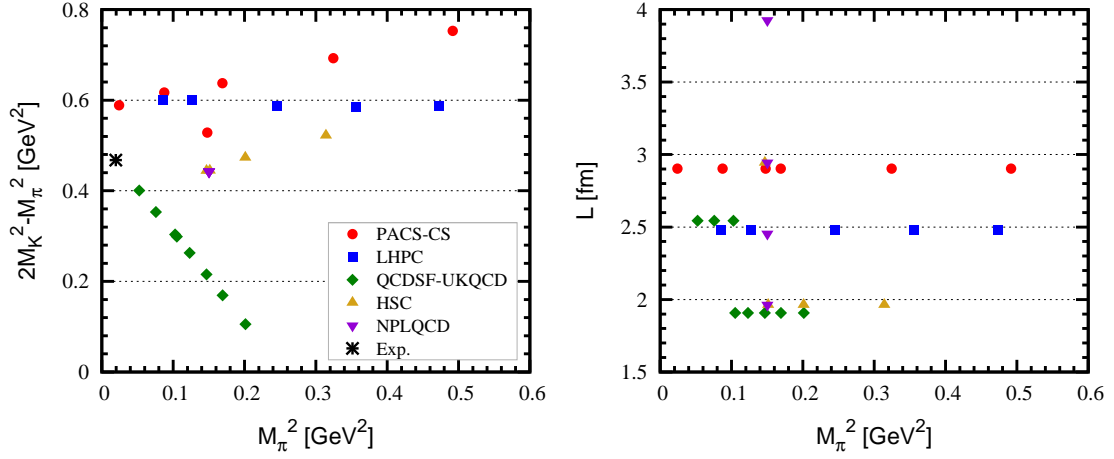}\caption{(Color online). Landscape of LQCD simulations of the ground-state octet baryon masses: the PACS-CS (red circles),  LHPC (blue squares), QCDSF-UKQCD (green diamonds), HSC (yellow upper triangles) and NPLQCD (purple lower triangles) collaborations in the $2M_K^2-M_{\pi}^2$ vs. $M_{\pi}^2$ plane (left panel) and in the $L$ vs. $M_{\pi}^2$ plane (right panel). The star denotes the physical point with the physical light- and strange-quark masses (as implied by leading order ChPT). Taken from Ref.~\cite{Ren:2012aj}.}
\label{fig:land}
\end{figure}

Up to N$^3$LO, the Feynman diagrams contributing to the octet baryon masses are shown in Fig.~\ref{dia:mass4}. They can be rather easily calculated and the detailed formulas can be found in Ref.~\cite{Ren:2012aj}.
There are in total 19 LECs that have to be determined, compared to only 4 at NNLO.

The LQCD simulation points are depicted in Fig.~\ref{fig:land} in the $M_\pi^2$ vs. $2M_K^2-M_\pi^2$ plane and in the $M_\pi^2$ vs. $L$ plane.
It is clear that to obtain the physical octet baryon masses, one has to extrapolate the LQCD simulation results to the physical light-quark masses and infinite space time. On the other hand,
as an perturbation theory up to a certain order, the covariant BChPT up to N$^3$LO is not expected to work up to arbitrarily large light-quark masses and arbitrarily small volumes. Therefore, one has  to select the LQCD simulations with a limited range of quark masses and volumes. In Ref.~\cite{Ren:2012aj}, 11 sets of LQCD data are selected with $M_\pi^2<0.5 $ GeV$^2$ and $M_\Phi L>4$ with $\Phi=\pi, K, \eta$. Fitting to the 11 sets of LQCD data plus the physical data, one obtains the results shown in Table \ref{table:mass4}.~\footnote{A second larger set of data, Set-II, has been used to demonstrate the applicability of covariant BChPT in
Ref.~\cite{Ren:2012aj}.} An order by order improvement of the description of the LQCD mass data is clearly seen from the decreasing $\chi^2/\mathrm{d.o.f}$ for each increasing chiral order.
The almost perfect fit at N$^3$LO indicates that the LQCD simulation results from different collaborations are consistent with each other. Furthermore, the values of the LECs at each order seem to be natural, i.e., the values are about $\mathcal{O}(1)$.

It should be stressed that a self-consistent consideration of finite-volume corrections is necessary to achieve a $\chi^2/.\mathrm{d.o.f}$ close to 1. Without taking into account finite-volume corrections,
 a $\chi^2/\mathrm{d.o.f}\approx1.9$ would have been obtained~\cite{Ren:2012aj}. In Ref.~\cite{Geng:2011wq},   it is shown that the NNLO EOMS BChPT can describe reasonably well the volume dependence of the NPLQCD data,
 which is consistent with the results of the N$^3$LO study performed in  Ref.~\cite{Ren:2012aj}.
\begin{table}[t]
\centering
\caption{Values of the LECs from the best fit to the LQCD data and the experimental data at $\mathcal{O}(p^2)$, $\mathcal{O}(p^3)$, and $\mathcal{O}(p^4)$ EOMS BChPT. Taken from Ref.~\cite{Ren:2012aj}.}
\label{table:mass4}
~~\\[0.1em]
\begin{tabular}{lrrr|r}
\hline\hline
         & \multicolumn{3}{c}{Set-I} & Set-II \\
      \cline{2-5}
         & Fit - $\mathcal{O}(p^2)$ & Fit - $\mathcal{O}(p^3)$  &   Fit I - $\mathcal{O}(p^4)$   &  Fit II - $\mathcal{O}(p^4)$  \\
\hline
  $m_0$~[MeV]         & $900(6)$      &   $767(6)$     &  $880(22)$      & $868(12)$     \\
  $b_0$~[GeV$^{-1}$]  &$-0.273(6)$    &  $-0.886(5)$   &  $-0.609(19)$   & $-0.714(21)$  \\
  $b_D$~[GeV$^{-1}$]  &$0.0506(17)$   &  $0.0482(17)$  &  $0.225(34)$    & $0.222(20)$   \\
  $b_F$~[GeV$^{-1}$]  &$-0.179(1)$    &  $-0.514(1)$   &  $-0.404(27)$   & $-0.428(12)$  \\
  $b_1$~[GeV$^{-1}$]  & --            &  --            &  $0.550(44)$    & $0.515(132)$   \\
    $b_2$~[GeV$^{-1}$]  & --            &  --            &  $-0.706(99)$   & $0.148(48)$   \\
  $b_3$~[GeV$^{-1}$]  & --            &  --            &  $-0.674(115)$  & $-0.663(155)$   \\
  $b_4$~[GeV$^{-1}$]  & --            &  --            &  $-0.843(81)$   & $-0.868(105)$    \\
  $b_5$~[GeV$^{-2}$]  & --            &  --            &  $-0.555(144)$  & $-0.643(246)$ \\
  $b_6$~[GeV$^{-2}$]  & --            &  --            &  $0.160(95)$    & $-0.268(334)$ \\
  $b_7$~[GeV$^{-2}$]  & --            &  --            &  $1.98(18)$     & $0.176(72)$  \\
  $b_8$~[GeV$^{-2}$]  & --            &  --            &  $0.473(65)$    & $-0.0694(1638)$ \\
  $d_1$~[GeV$^{-3}$]  & --            &  --            &  $0.0340(143)$  & $0.0345(134)$  \\
  $d_2$~[GeV$^{-3}$]  & --            &  --            &  $0.296(53)$    & $0.374(21)$    \\
  $d_3$~[GeV$^{-3}$]  & --            &  --            &  $0.0431(304)$  & $0.00499(1817)$ \\
  $d_4$~[GeV$^{-3}$]  & --            &  --            &  $0.234(67)$    & $0.267(34)$     \\
  $d_5$~[GeV$^{-3}$]  & --            &  --            &  $-0.328(60)$   & $-0.445(26)$    \\
  $d_7$~[GeV$^{-3}$]  & --            &  --            &  $-0.0358(269)$ & $-0.183(12)$    \\
  $d_8$~[GeV$^{-3}$]  & --            &  --            &  $-0.107(32)$   & $-0.307(21)$    \\
  \hline
$\chi^2$/d.o.f. & $11.8$ & $8.6$  & $1.0$ & $1.6$\\
\hline\hline
\end{tabular}
\end{table}
 \subsubsection{Sigma terms of the octet baryons}
A natural application of the above study is to use the so-obtained LECs to predict the baryon sigma terms, which play an important role in understanding chiral symmetry breaking and in direct dark matter searches~\cite{Bottino:1999ei,Ellis:2008hf}.
 They are related to the light-quark mass dependence of the
baryon masses via the Feynman-Hellmann theorem, which states:
\begin{eqnarray}\label{sigmacal}
  \sigma_{\pi B} &=& m_l\langle B(p)|\bar{u}u+\bar{d}d|B(p)\rangle = m_{l}\frac{\partial M_B}{\partial m_l},\\
  \sigma_{s B} &=& m_s\langle B(p)|\bar{s}s|B(p)\rangle = m_s\frac{\partial M_B}{\partial m_s},
\end{eqnarray}
where $m_l=(m_u+m_d)/2$. Using the leading order meson ChPT, the quark masses can be expressed by the pseudoscalar masses, with $m_l=M_{\pi}^2/(2B_0)$ and $m_s=(2M_K^2-M_{\pi}^2)/(2B_0)$.
Other related quantities, which often appear in the literature, including the strangeness content ($y_B$) and the so-called ``dimensionless sigma terms" ($f_{lB}$, $f_{sB}$), can be  calculated as well:
\begin{eqnarray}\label{strangenesscal}
  y_B &=& \frac{2\langle B(p)|\bar{s}s|B(p)\rangle}{\langle B(p)|\bar{u}u+\bar{d}d|B(p)\rangle}=\frac{m_l}{m_s}\frac{2\sigma_{sB}}{\sigma_{\pi B}},\\
  f_{lB} &=& \frac{m_{l}\langle B(p)|\bar{u}u+\bar{d}d|B(p)\rangle}{M_B}=\frac{\sigma_{\pi B}}{M_B},\\
  f_{sB} &=& \frac{m_s\langle B(p)|\bar{s}s|B(p)\rangle}{M_B}=\frac{\sigma_{s B}}{M_B}.
\end{eqnarray}

Using the $\mathcal{O}(p^4)$ Fit-I low-energy constants tabulated in Table \ref{table:mass4},  the pion- and strangeness-sigma terms $\sigma_{\pi B}$, $\sigma_{s B}$ for all the  octet baryons, and the corresponding strangeness content $y_{B}$, ``dimensionless sigma terms" $f_{lB}$, $f_{sB}$ can be calculated and are shown in Table \ref{sigmaresults}. The nucleon pion-sigma term at the physical point, $\sigma_{\pi N}=43(1)(6)$ MeV, is in reasonable agreement with the determination in the study of the old $\pi-N$ scattering data~\cite{Gasser:1990ce} of $\sigma_{\pi N}=45\pm 8$ MeV, but smaller than the central value of the more recent study,  $\sigma_{\pi N}=59\pm 7$ MeV~\cite{Alarcon:2011zs}. The $\sigma_{\pi N}$ is  also in agreement with the recent lattice result of the QCDSF collaboration ($\sigma_{\pi N}=38(12)$ MeV~\cite{Bali:2011ks}, $\sigma_{\pi N}=37(8)(6)$ MeV~\cite{Bali:2012qs}) and BMW collaboration ($\sigma_{\pi N}=39(4)^{+18}_{-7}$~MeV)~\cite{Durr:2011mp} within uncertainties, and is consistent with the HBChPT result of Ref.~\cite{Shanahan:2012wh}. On the other hand, it is larger than the QCDSF result ($\sigma_{\pi N}=31(3)(4)$ MeV)~\cite{Horsley:2011wr} and the N$^3$LO BChPT results using the partial summation scheme ($\sigma_{\pi N}=32(1)$ MeV)~\cite{Semke:2012gs}, but slightly smaller than the JLQCD result ($\sigma_{\pi N}=50(4.5)$ MeV)~\cite{Ohki:2009mt}.

It should be mentioned that the central value of the nucleon sigma term $\sigma_{\pi N}$  is smaller than that of Ref.~\cite{MartinCamalich:2010fp},
where the result is obtained with the NNLO EOMS BChPT by fitting the PACS-CS data. Inclusion of virtual decuplet contributions may have some non-negligible effects on the predicted baryon sigma terms. At NNLO, it is found that the inclusion of virtual decuplet baryons can increase the pion-nucleon sigma term while decrease the strangeness-nucleon sigma term~\cite{Alarcon:2012nr}. It is interesting to check whether such effects still exist at N$^3$LO.

The predicted strangeness-nucleon sigma term is larger than those predicted in Refs.~\cite{Shanahan:2012wh,Semke:2012gs}. It will be interesting to find out the origin of such discrepancies by studying the dependence on different formulations of BChPT and on LQCD data.~\footnote{For a related discussion, see Ref.~\cite{Shanahan:2013cd}.}

\begin{table}[t]
\centering
\caption{The sigma-terms, the strangeness content, and the ``dimensionless sigma terms" of the octet baryons at the physical point.
The first error is statistical and the second one is systematic, estimated by taking half the difference between the N$^3$LO result and the NNLO result.
Taken from Ref.~\cite{Ren:2012aj}.}
~~\\[0.1em]
\begin{tabular}{cccccccc}
\hline\hline
       & $\sigma_{\pi B}$~[MeV]  & $\sigma_{sB}$~[MeV]  & $y_{B}$  & $f_{lB}$  & $f_{sB}$ \\
\hline
$N$       &  $43(1)(6)$  & $126(24)(54)$    &  $0.244(47)(110)$    & $0.0457(11)(64)$   & $0.134(26)(57)$ \\
$\Lambda$ &  $19(1)(7)$  & $269(23)(66)$    &  $1.179(118)(522)$   & $0.0170(9)(63)$   & $0.241(21)(59)$ \\
$\Sigma$  &  $18(2)(6)$  & $296(21)(50)$    &  $1.369(180)(512)$   & $0.0151(17)(50)$   & $0.248(18)(42)$ \\
$\Xi$     &  $ 4(2)(3)$  & $397(22)(56)$    &  $8.263(4157)(6306)$ & $0.00303(152)(228)$ & $0.301(17)(42)$ \\
\hline\hline
\end{tabular}
\label{sigmaresults}
\end{table}

\subsection{Hyperon vector couplings}

Hyperon semileptonic decays, parameterized by three vector transition form factors ($f_1$, $f_2$, and $f_3$) and three axial form factors ($g_1$, $g_2$, and $g_3$), have received renewed interest in
recent years due to various reasons. In particular,
they provide an alternative source~\cite{Cabibbo:2003cu,Cabibbo:2003ea,FloresMendieta:2004sk,Mateu:2005wi}
to allow one to extract the Cabibbo-Kobayashi-Maskawa (CKM)
matrix element $V_{us}$ ~\cite{Cabibbo:1963yz,Kobayashi:1973fv},
in addition to kaon semileptonic decays (see, e.g., Ref.~\cite{Blucher:2005dc} for a recent review),
hadronic decays of the $\tau$ lepton~\cite{Gamiz:2007qs} and the ratio $\Gamma(K^+\rightarrow \mu^+\nu_\mu)/\Gamma(\pi^+\rightarrow\mu^+\nu_\mu)$~\cite{Marciano:2004uf}.
The hyperon vector coupling $f_1(0)$ plays an essential role in order to extract $V_{us}$ accurately.

Because of the Conservation of Vector Current (CVC), $f_1(0)$ is
known up to SU(3)-breaking effects, which are of subleading-order according to the Ademollo-Gatto theorem~\cite{Ademollo:1964sr}. Theoretical estimates of SU(3)-breaking corrections to $f_1(0)$ have
been performed in various frameworks, including quark models~\cite{Donoghue:1986th,Schlumpf:1994fb,Faessler:2008ix},
 large-$N_c$ fits~\cite{FloresMendieta:2004sk}, and
BChPT~\cite{Krause:1990xc,Anderson:1993as,Kaiser:2001yc,Villadoro:2006nj,Lacour:2007wm,Geng:2009ik}.
These SU(3)-breaking corrections have also been studied recently in quenched LQCD simulations \cite{Guadagnoli:2006gj,Sasaki:2008ha}, and
more lately in $n_f=2+1$ LQCD simulations~\cite{Sasaki:2012ne}.

In principle, ChPT provides a model independent way to
estimate the SU(3)-breaking corrections to $f_1(0)$.  However, it is known that ChPT calculations
converge slowly in the three-flavor sector of $u$, $d$, and $s$ quarks. This problem becomes even more pronounced
in the one-baryon sector, where the physics at a certain order can be blurred
by the power-counting restoration procedures, as can be clearly seen in
the case of the baryon octet magnetic moments~\cite{Geng:2008mf}.
Fortunately, in the case of  $f_1(0)$,
the Ademollo-Gatto theorem dictates that up to $\mathcal{O}(p^4)$ no unknown LECs contribute and, therefore, no
power-counting-breaking terms appear. Consequently, up to this order there is no need to apply any
power-counting restoration procedures and
a BChPT calculation is fully predictive.

In a recent $\mathcal{O}(p^4)$ calculation performed in the heavy-baryon ChPT~\cite{Villadoro:2006nj}, it was shown that the chiral series with only the octet contributions converge slowly  while the convergence is
completely spoiled by the inclusion of the decuplet ones.
In a later work~\cite{Lacour:2007wm}, the infrared BChPT was employed and
calculations were performed up to $\mathcal{O}(p^4)$ with only
the octet contributions. The slow convergence of the chiral series was confirmed but the importance of relativistic corrections was stressed. The first $\mathcal{O}(p^4)$ EOMS BChPT study was performed in Ref.~\cite{Geng:2009ik}, including the contributions of both the virtual
octet and the virtual decuplet baryons.

The baryon vector form factors as probed by the charged $\Delta$S=1 weak current
$V^\mu=V_{us}\bar{u}\gamma^\mu s$
are defined by
\begin{equation}
\langle B'\vert V^\mu\vert B\rangle =V_{us}\bar{u}(p')\left[\gamma^\mu f_1(q^2)+\frac{2i \sigma^{\mu\nu}q_\nu}{M_{B'}+M_B}f_2(q^2)+\frac{2 q^\mu }{M_{B'}+M_B}f_3(q^2)\right]u(p),
\end{equation}
where $q=p'-p$. In the SU(3)-symmetric limit, $f_1(0)$ is fixed by the conservation of the SU(3)$_V$-charge $g_V$. Furthermore, the Ademollo-Gatto theorem states that  SU(3)-breaking corrections start at second order in the expansion parameter $m_s-m$
\begin{equation}\label{eq:ag}
f_1(0)=g_V+\mathcal{O}((m_s-m)^2),
\end{equation}
where $m_s$ is the strange quark mass and $m$ is the average mass of the light quarks.
The values of $g_V$ are $-\sqrt{\frac{3}{2}}$, $-\frac{1}{\sqrt{2}}$, $-1$,  $\sqrt{\frac{3}{2}}$, $\frac{1}{\sqrt{2}}$,  $1$ for  $\Lambda\rightarrow p$, $\Sigma^0\rightarrow p$,
 $\Sigma^-\rightarrow n$,  $\Xi^-\rightarrow \Lambda$, $\Xi^-\rightarrow \Sigma^0$, and
$\Xi^0\rightarrow\Sigma^+$, respectively.
In the isospin-symmetric limit only four of these channels, which are taken as $\Lambda\rightarrow N$, $\Sigma\rightarrow N$, $\Xi\rightarrow\Lambda$, and $\Xi\rightarrow\Sigma$, provide independent information.  The SU(3)-breaking corrections are parameterised order-by-order in the relativistic chiral expansion as follows:
\begin{equation}
f_1(0)=g_V\left( 1+\delta^{(2)}+\delta^{(3)}+\cdots\right) \label{eq:Adem-Gatt},
\end{equation}
where $\delta^{(2)}$ and $\delta^{(3)}$ are the leading order  and next-to-leading order SU(3)-breaking corrections
induced by loops, corresponding to $\mathcal{O}(p^3)$ and $\mathcal{O}(p^4)$ chiral calculations.

\begin{figure}[t]
\includegraphics[scale=0.8,angle=270]{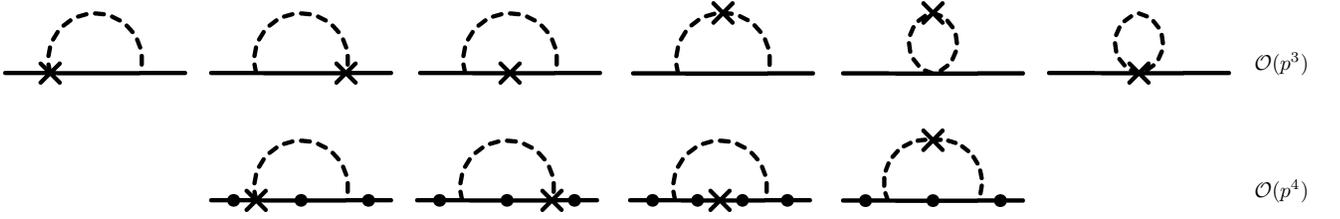}
\caption{Feynman diagrams contributing to the SU(3)-breaking corrections to
the hyperon vector coupling $f_1(0)$
up to $\mathcal{O}(p^4)$. The solid lines correspond to baryons and dashed lines to mesons; crosses indicate the coupling of the external current; black dots denote mass splitting
insertions. The diagrams corresponding to wave function renormalization are not explicitly shown but are taken into account in the calculation.
\label{fig:doctet}}
\end{figure}
\begin{figure}[t]
\includegraphics[scale=0.8,angle=270]{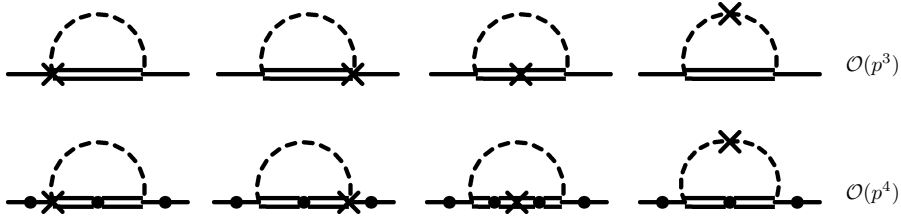}
\caption{Feynman diagrams contributing to the LO and NLO SU(3)-breaking corrections
to the hyperon vector coupling $f_1(0)$, through dynamical decuplet baryons. The
notations are the same as those of Fig.~\ref{fig:doctet} except that double lines
indicate decuplet baryons. \label{fig:ddecup}}
\end{figure}
\begin{table}[htbp]
      \renewcommand{\arraystretch}{1.5}
     \setlength{\tabcolsep}{0.4cm}
     \caption{Values for the masses and couplings appearing in the calculation of
the SU(3)-breaking corrections to $f_1(0)$. \label{table:para}}
\begin{tabular}{cl|cl}
\hline\hline
 $D$& $0.8$ & $M_B$ & $1.151$ GeV \\
 $F$& $0.46$ & $M_D$ &$1.382$ GeV\\
 $f_\pi$ & $0.0924$ GeV & $M_0$ & $1.197$ GeV \\
 $F_0$& $1.17f_\pi$ & $b_D$ & $-0.0661$ GeV$^{-1}$\\
 $m_\pi$ & $0.138$  GeV&$b_F$ & $0.2087$ GeV$^{-1}$\\
 $m_K$ & $0.496$  GeV & $M_{D0}$ & $1.216$ GeV \\
 $m_\eta$ & $0.548$ GeV & $\gamma_M$ & $0.3236$ GeV$^{-1}$\\
 $\mathcal{C}$& 1.0\\
\hline\hline
\end{tabular}
\end{table}

Up to $\mathcal{O}(p^4)$, one has to calculate the diagrams shown in Figs.~\ref{fig:doctet} and \ref{fig:ddecup}.  Using the LECs given in Table \ref{table:para}, one can calculate the contributions of virtual octet baryons
and virtual decuplet baryons.  The virtual octet contributions  are shown in Table \ref{table:octetfull} in comparison with the heavy-baryon and infrared BChPT results, when available.  It is seen that
the heavy-baryon results are qualitatively different from those of the IR and EOMS results at $\mathcal{O}(p^4)$. In three of the four cases, the sign of $\delta(2)+\delta(3)$ is different. On the other hand, the EOMS and IR results
agree much better with each other. Furthermore, one notices that even in the EOMS case, convergence seems to be slow.
The contributions of virtual decuplet baryons are given in Table \ref{table:decupfull}, in comparison with the HB results. One notes that in general the contributions of the virtual decuplet baryons are
not negligible and their contributions to the SU(3) breaking corrections are positive.

In Table~\ref{table:comparison}, the complete $\mathcal{O}(p^4)$ covariant BChPT predictions are compared with the results of a number of other approaches.
They seem to agree with those of the chiral quark model~\cite{Faessler:2008ix}  and the large $N_c$~\cite{FloresMendieta:2004sk} results, but are only partially consistent with  the quenched LQCD simulations~\cite{Guadagnoli:2006gj,Sasaki:2008ha}.

It is surprising that, as also pointed out in Ref.~\cite{Sasaki:2012ne}, the tendency of the SU(3) breaking correction observed in the $n_f=2+1$ LQCD simulations
 disagrees with predictions of both the latest BChPT result~\cite{Geng:2009ik}  and the large $N_c$ analysis~\cite{FloresMendieta:2004sk}  . It will be of great importance to find out  how to reconcile the BChPT results
 with the dynamical LQCD simulations. A careful study of the quark-mass, lattice-volume, and even finite lattice-spacing  dependence of the LQCD simulations may have to be performed in order to understand the
 observed discrepancy~\cite{Sasaki:2012ne}.

\begin{table}[t]
      \renewcommand{\arraystretch}{1.5}
     \setlength{\tabcolsep}{0.4cm}
     \caption{Octet contributions to the SU(3)-breaking corrections to $f_1(0)$ (in percentage). The central values of the  $\mathcal{O}(p^4)$ results are calculated with $\mu=1$ GeV and the uncertainties are
obtained by varying $\mu$ from 0.7 to 1.3 GeV. Taken from Ref.~\cite{Geng:2009ik}.\label{table:octetfull}}
\begin{tabular}{c|ccc|c|c}
\hline\hline
&\multicolumn{3}{c|} {present work}& HBChPT~\cite{Villadoro:2006nj} & IRChPT~\cite{Lacour:2007wm}\\\hline
                        & $\delta^{(2)}$&  $\delta^{(3)}$ & $\delta^{(2)}+\delta^{(3)}$& $\delta^{(2)}+\delta^{(3)}$& $\delta^{(2)}+\delta^{(3)}$\\
			\hline
 $\Lambda\rightarrow N$ & $-3.8$  &  $0.2^{+1.2}_{-0.9}$ & $-3.6^{+1.2}_{-0.9}$  & $2.7$ & $-5.7\pm2.1$ \\
 $\Sigma\rightarrow N$  & $-0.8$ &  $4.7^{+3.8}_{-2.8}$ & $3.9^{+3.8}_{-2.8}$   & $4.1$ & $2.8\pm0.2$\\
 $\Xi\rightarrow\Lambda$& $-2.9$  &  $1.7^{+2.4}_{-1.8}$ & $-1.2^{+2.4}_{-1.8}$ & $4.3$ & $-1.1\pm1.7$\\
 $\Xi\rightarrow\Sigma$ & $-3.7$  &  $-1.3^{+0.3}_{-0.2}$ & $-5.0^{+0.3}_{-0.2}$  & $0.9$ & $-5.6\pm1.6$\\
 \hline\hline
\end{tabular}
\end{table}

\begin{table}[t]
      \renewcommand{\arraystretch}{1.5}
     \setlength{\tabcolsep}{0.4cm}
     \caption{Decuplet contributions to the SU(3)-breaking corrections to $f_1(0)$ (in percentage). The central values of the $\mathcal{O}(p^4)$ result are calculated with $\mu=1$ GeV and the uncertainties are obtained by varying $\mu$ from 0.7 to 1.3 GeV.
     Taken from Ref.~\cite{Geng:2009ik}.
     \label{table:decupfull}}
\begin{tabular}{c|ccc|ccc}
\hline\hline
&\multicolumn{3}{c}{Present work}&\multicolumn{3}{c}{HBChPT}\\\hline
& $\delta^{(2)}$&  $\delta^{(3)}$ & $\delta^{(2)}+\delta^{(3)}$  & $\delta^{(2)}$&  $\delta^{(3)}$ & $\delta^{(2)}+\delta^{(3)}$\\
			\hline
 $\Lambda\rightarrow N$ &    $0.7$ & $3.0^{+0.1}_{-0.1}$ & $3.7^{+0.1}_{-0.1}$ & $1.8$ & $1.3$ & $3.1$ \\
 $\Sigma\rightarrow N$  & $-1.4$ & $6.2^{+0.4}_{-0.3}$ & $4.8^{+0.4}_{-0.3}$& $-3.6$ & $8.8$ & $5.2$\\
 $\Xi\rightarrow\Lambda$&  $-0.02$ & $5.2^{+0.4}_{-0.3}$ & $5.2^{+0.4}_{-0.3}$ &$-0.05$ & $4.2$ & $4.1$ \\
 $\Xi\rightarrow\Sigma$ &  $0.7$ & $6.0^{+1.9}_{-1.4}$ & $6.7^{+1.9}_{-1.4}$ &$1.9$ & $-0.2$ & $1.7$\\
 \hline\hline
\end{tabular}
\end{table}

\begin{table}[b]
      \renewcommand{\arraystretch}{1.}
     \setlength{\tabcolsep}{0.4cm}
     \caption{SU(3)-breaking corrections (in percentage) to $f_1(0)$ obtained
in different approaches.
     \label{table:comparison}}
\begin{tabular}{c|c|c|ccc|cc}
\hline\hline
& EOMS BChPT~\cite{Geng:2009ik} & Large $N_c$ & \multicolumn{3}{c|}{Quark model} &LQCD \\\hline
& &Ref.~\cite{FloresMendieta:2004sk}   & Ref.~\cite{Donoghue:1986th} &Ref.~\cite{Schlumpf:1994fb} & Ref.~\cite{Faessler:2008ix} &  \\\hline
\footnotesize
 $\Lambda\rightarrow N$ &    $0.1^{+1.3}_{-1.0}$ & $2\pm2$ & $-1.3$ & $-2.4$ & $0.1$ & \\
 $\Sigma\rightarrow N$  & $8.7^{+4.2}_{-3.1}$ &  $4\pm3$ &  $-1.3$ & $-2.4$ & $0.9$ & $-1.2\pm2.9\pm4.0$~\cite{Guadagnoli:2006gj}\\
                                           &                                     &                    &              &             &             & $-3.02\pm1.06\pm0.15\pm0.36$~\cite{Sasaki:2012ne}\\
 $\Xi\rightarrow\Lambda$ &  $4.0^{+2.8}_{-2.1}$&  $4\pm4$ &  $-1.3$ & $-2.4$ & $2.2$ & \\
 $\Xi\rightarrow\Sigma$ &  $1.7^{+2.2}_{-1.6}$ &  $8\pm5$ &  $-1.3$ & $-2.4$ & $4.2$ & $-1.3\pm1.9$~\cite{Sasaki:2008ha}\\
 &                                        &                                           &              &              &              &             $-2.68\pm0.66\pm0.07\pm0.05$~\cite{Sasaki:2012ne}\\
 \hline\hline
\end{tabular}
\end{table}

\section{Recent applications in the $u$, $d$ two-flavor sector}
Although in this short review we mainly focus on the $u$, $d$, $s$ three-flavor sector, we must point out that there are many interesting and important developments in the two-flavor sector. For instance, the pion-nucleon scattering has been studied up to $\mathcal{O}(p^3)$ in Refs.~\cite{Alarcon:2011zs,Alarcon:2012kn} and up to $\mathcal{O}(p^4)$ in Ref.~\cite{Chen:2012nx}. It was shown that  the EOMS approach overcomes the limitations that previous chiral analyses of the $\pi N$ scattering amplitude had, and provides an accurate description of the partial wave phase shifts of the Karlsruhe-Helsinki and George-Washington groups up to energies just below the resonance region. Furthermore, the EOMS BChPT is shown to exhibit a good convergence~\cite{Chen:2012nx}. The proton Compton scattering has been studied in Ref.~\cite{Lensky:2009uv} and the nucleon and $\Delta(1232)$-isobar electromagnetic form factors in Ref.~\cite{Ledwig:2011cx}.  The nucleon to $\Delta$ axial transition form factors have been studied in Ref.~\cite{Geng:2008bm}.
The nucleon mass has been studied up to  $\mathcal{O}(p^6)$~\cite{Schindler:2006ha,Schindler:2007dr}, the axial, induced pseudoscalar, and pion-nucleon form factors up to $\mathcal{O}(p^4)$~\cite{Schindler:2006it}, and the electromagnetic form factors of the nucleon up to $\mathcal{O}(p^4)$~\cite{Schindler:2005ke}.
The amplitude for ordinary muon capture on the proton has been studied as well~\cite{Ando:2006xy}.

Some of these studies showed clear improvement of the EOMS scheme over the IR or HB formulations (see, e.g., Refs.~\cite{Alarcon:2011zs,Ledwig:2011cx}), while others did not (see, e.g., Ref.~\cite{Ando:2006xy}).  Nevertheless, it is always preferable  to perform a study with a fully covariant ChPT supplemented with a power-counting restoration procedure that does not
spoil the analytical structure of the loop results, e.g., the EOMS BChPT.

There is no doubt that in principle
the EOMS BChPT should converge faster/better in the two-flavor sector than in the three-flavor sector, because of the large expansion parameter $M_K/\Lambda_{\chi \mathrm{PT}}$ in the latter case.
Nevertheless, in the last few years, it has been shown that the EOMS BChPT also converges as expected in the three-flavor sector, as has been reviewed in this work. In future, 
one may wish to study some observables with and without the strange quark degree of freedom integrated out and then relate the relevant LECs in the two-flavor sector and
those in the three-flavor sector. Such studies may help to better constrain  the large number of unknown/poorly known LECs in the three-flavor sector.
We note that such a study of meson-baryon scattering lengths has been performed in the infrared BChPT~\cite{Mai:2009ce}.

\section{Extensions to the heavy-light system}
\begin{figure}[t]
\centerline{\includegraphics[scale=1.0]{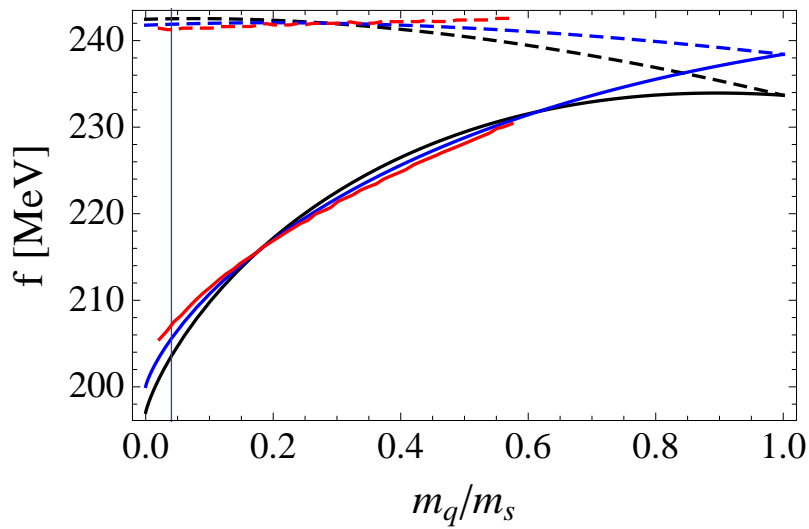}}
\caption{(Color online) Light-quark mass dependence of $f_D$  (solid lines) and $f_{D_s}$ (dashed lines).
The black lines show the results of the NLO HM$\chi$PT and the blue lines the results of the covariant NLO ChPT.
The red lines are the continuum extrapolations of the HPQCD collaboration~\cite{Follana:2007uv}.
The ratio $r=m_q/m_s$ is related to the pseudoscalar meson masses at leading chiral order through
$m_\pi^2=2B_0 m_s r$ and $m_K^2=B_0 m_s (r+1)$ with $B_0=m_\pi^2/(m_u+m_d)$  and
$m_q=(m_u+m_d)/2$, where $m_u$, $m_d$, and $m_s$ are the
physical up, down, and strange quark mass. Taken from Ref.~\cite{Geng:2010df}.
\label{fig:fdfds}}
\end{figure}

Given the formal similarity between the heavy-light mesons and the ground-state baryons, in particular, regarding chiral descriptions of their static properties and
scattering with the NGBs,  the covariant formulation of ChPT has been extended to study the decay constants of $D$ and $B$ mesons~\cite{Geng:2010df,Altenbuchinger:2011qn}, and
the scattering lengths of the NGBs off the $D$ mesons~\cite{Geng:2010vw}. Indeed, the covariant formulation also shows improved convergence in comparison with the
conventional heavy-meson (HM) ChPT.  From Fig.~\ref{fig:fdfds}, it is clear that with the same number of LECs, the covariant ChPT describes better the light-quark mass dependence of
the LQCD results than the non-relativistic HM ChPT.

 Recently, the infrared formulation of ChPT has been applied to study
the pseudoscalar meson and heavy vector meson scattering lengths as well~\cite{Liu:2011mi}.

\section{Summary and Outlook}
In this work, we have briefly reviewed some recent developments in covariant baryon chiral perturbation theory (BChPT). After a short introduction to the power-counting-breaking problem appearing in the formulation of BChPT and the existing
recipes to recover a systematic power-counting, we have summarized the advantages and limitations of each method and have shown a number of successful applications of the EOMS scheme in the
$u$, $d$, and $s$ three-flavor sector, including the magnetic moments of the octet and decuplet baryons, the masses and sigma terms of the octet baryons, and the hyperon vector couplings. The developments in
the $u$ and $d$ two-flavor sector, as well as the recent extensions to the heavy-light system, are also briefly covered.

It is shown that the covariant BChPT descriptions of the magnetic moments of the octet and decuplet baryons and the masses of the octet baryons are very satisfactory, but
for the strangeness-nucleon sigma term and the hyperon vector couplings, more works are still needed  to reconcile the BChPT predictions with the LQCD data or to understand the large discrepancy between different studies within the BChPT framework.
In this regard, effects of higher-energy degrees of freedom have to be examined more carefully, e.g., the decuplet baryons.
 Although virtual decuplet contributions are found to be small
in the case of the magnetic moments of the octet baryons, their contributions to the hyperon vector couplings are not negligible. As having been shown in the heavy-baryon ChPT,
one may have to study the effects of virtual decuplet baryons case by case.
Furthermore, the extension to the heavy-light system should be further explored.

Compared to its non-relativistic and infrared counterparts, the EOMS formulation seems to be more suitable for the three-flavor sector and for descriptions of LQCD simulation data.
One should note, however, that a systematic study of different physical observables in the three-flavor sector using the EOMS formulation is still missing. Such a study is necessary not only for cross-checking the validity of the EOMS BChPT but also for  truly testing the predictive power of the covariant BChPT.  In view of the rapid progress being made in both experiments and LQCD simulations, such studies are strongly encouraged.

The manifestly Lorentz covariant formulation of BChPT can also be applied
to study baryon-baryon scattering.  The investigations in this direction  could
further deepen our understanding of the nucleon-nucleon, hyperon-nucleon, and hyperon-hyperon forces. With
the relevant low-energy constants determined either with the help of  LQCD simulations or through fitting experimental data,
the baryon-baryon forces can be used in  studies of the properties of atomic nuclei/hypernuclei and dense stellar objects, such as neutron stars.

\section{Acknowledgements}
L. S. G. acknowledges fruitful discussions
with L. Alvarez-Ruso, M. Altenbuchinger, N. Kaiser,  J. Martin-Camalich, J. Meng,  X.-L. Ren, H. Toki, M.J. Vicente Vacas,  and W. Weise.
This work was partly supported by the National Natural Science Foundation of China under Grants No. 11005007, No. 11035007, and No. 11175002,
and the New Century Excellent Talents in University Program of Ministry of Education of China under Grant No. NCET- 10-0029.

\bibliography{cbchpt-ref}{}

\end{document}